\documentclass[10pt,acmsmall, nonacm]{acmart}
\renewcommand\footnotetextcopyrightpermission[1]{}

\makeatletter
\def\@ACM@checkaffil{
    \if@ACM@instpresent\else
    \ClassWarningNoLine{\@classname}{No institution present for an affiliation}%
    \fi
    \if@ACM@citypresent\else
    \ClassWarningNoLine{\@classname}{No city present for an affiliation}%
    \fi
    \if@ACM@countrypresent\else
        \ClassWarningNoLine{\@classname}{No country present for an affiliation}%
    \fi
}
\makeatother

\usepackage{url}
\usepackage{enumerate}
\usepackage{subfigure}
\usepackage{multirow}
\usepackage{footnote}
\usepackage{verbatim}
\usepackage{booktabs}
\usepackage{threeparttable}
\usepackage{multicol} 
\usepackage{color}
\usepackage{color}
\usepackage{transparent}
\usepackage{import}
\usepackage{svg}
\usepackage{longtable}
\usepackage{xtab}
\usepackage{pifont}
\usepackage{enumitem}
\usepackage{cuted}
\usepackage{bbding}
\usepackage{listings}
\usepackage[mathscr]{eucal}
\usepackage[utf8]{inputenc}
\usepackage{cleveref}
\usepackage{makecell}

\crefname{section}{§}{§§}
\Crefname{section}{§}{§§}
\usepackage{soul}

\acmBooktitle{}
\begin{document}
\title{Do NFTs' Owners Really Possess their Assets? \\ A First Look at the NFT-to-Asset Connection Fragility}

\author[1]{Ziwei Wang}
\affiliation{Southern University of Science and Technology}
\author[1]{Jiashi Gao}
\affiliation{Southern University of Science and Technology}
\author[1]{Xuetao Wei \thanks{}}
\affiliation{Southern University of Science and Technology}

\makeatletter
\let\@authorsaddresses\@empty
\makeatother

\begin{abstract}

NFTs (Non-Fungible Tokens) have experienced an explosive growth and their record-breaking prices have been witnessed. Typically, the assets that NFTs represent are stored off-chain with a pointer, \textit{e.g.}, multi-hop URLs, due to the costly on-chain storage. Hence, this paper aims to answer the question: Is the NFT-to-Asset connection fragile? This paper makes a first step towards this end by characterizing NFT-to-Asset connections of 12,353 Ethereum NFT Contracts (6,234,141 NFTs in total) from three perspectives, storage, accessibility and duplication. In order to overcome challenges of affecting the measurement accuracy, \textit{e.g.}, IPFS instability and the changing availability of both IPFS and servers' data, we propose to leverage multiple gateways to enlarge the data coverage and extend a longer measurement period with non-trivial efforts. Results of our extensive study show that such connection is very fragile in practice. The loss, unavailability, or duplication of off-chain assets could render value of NFTs worthless. For instance, we find that assets of 25.24\% of Ethereum NFT contracts are not accessible, and 21.48\% of Ethereum NFT contracts include duplicated assets. Our work sheds light on the fragility along the NFT-to-Asset connection, which could help the NFT community to better enhance the trust of off-chain assets.


\end{abstract}


\maketitle

\section{Introduction}
\label{Introduction}
    The year of 2021 is remarkable for NFTs since the market of NFTs has been expanded to \$7B. The record-breaking prices have been witnessed, such as the ``Everydays: The First 5,000 Days'' from Beeple \cite{Beeple} was sold at \$69.3M and ``The Merge'' from Pak was sold at \$91.8M \cite{Merge}. The value of an NFT largely comes from the digital asset it represents, as well as the unique and identifiable ownership of such asset backed by blockchain. 

\begin{figure*}[htbp] 
            \centering
            \includegraphics[width=0.8\textwidth]{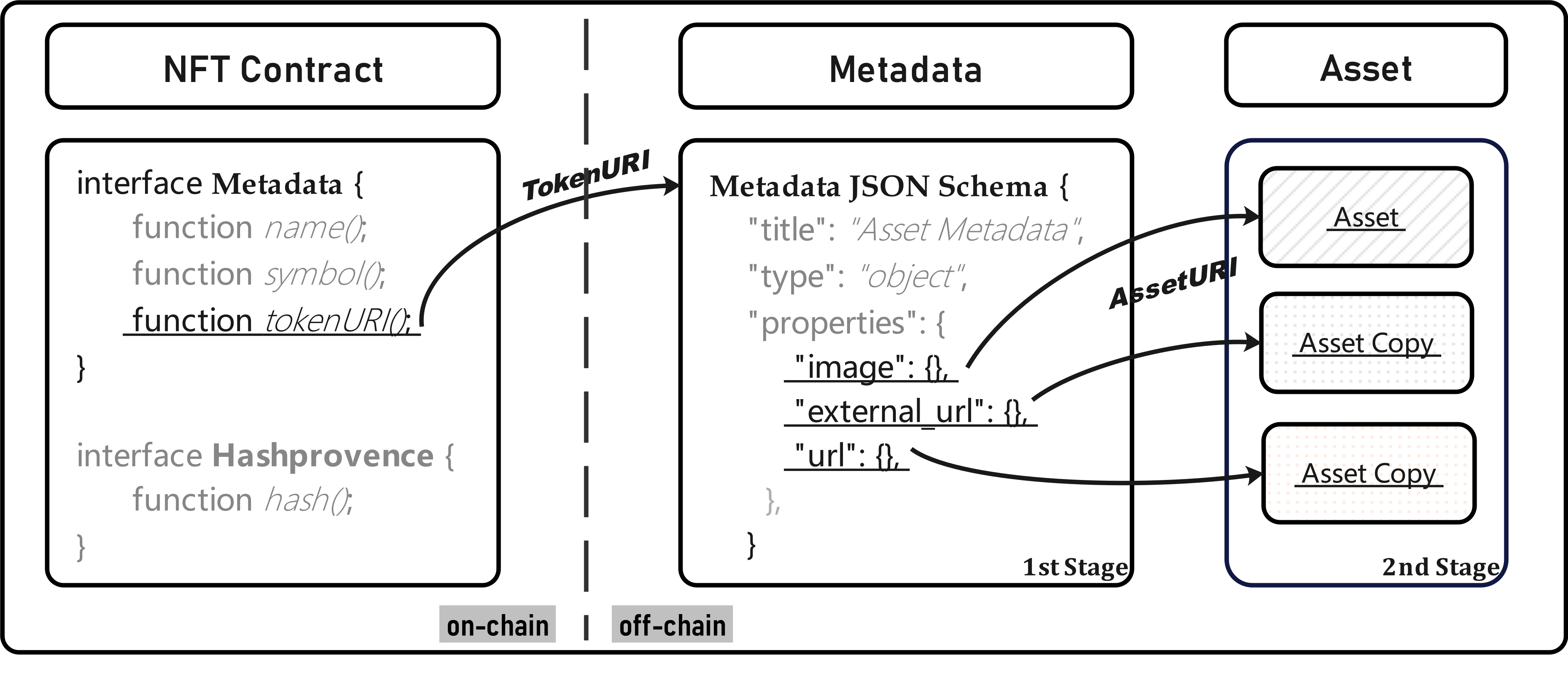}
            \caption{The anatomy of the NFT-to-Asset connection.}
            \label{nft2asset}
        \end{figure*}

    Due to the costly storage on the blockchain (on-chain), most of NFTs choose to save their assets off the blockchain (off-chain). Centralized storage, \textit{i.e.}, storing in NFT providers' own servers or cloud services, and decentralized storage, \textit{e.g.}, \textit{InterPlanetary File System} (\textsc{IPFS}) \cite{IPFS}, \textsc{Filecoin} \cite{Filecoin}, \textsc{Arweaves} \cite{Arweaves}, are used to host these assets. Decentralized storage is generally considered with better sustainability and tamper-resistance as data is not controlled by a centralized entity. There are two ways to bind the on-chain NFT with the off-chain asset: \ding{182} the metadata field in NFTs records multi-stage storage addresses to point to the off-chain asset, as shown in Figure~\ref{nft2asset}, which refers to the NFT-to-Asset connection in this study; \ding{183} a hash value or the Merkle root of a group of data is recorded in the metadata of NFTs, which is not widely adopted in practice. Note that in this work, we focus on the asset path of NFTs and won't discuss the effect of hash proof on the asset safety since the NFTs' hashes are unable to be retrieved with standardized \texttt{ABI}. It will be a future research area to combine cryptography with the NFT to enhance NFT assets' safety. 

    Previous work has investigated NFTs on price forecasting \cite{nadini2021mapping, jain2022nft}, integrity \cite{wang2021non, das2021understanding}, plagiarism \cite{pungila2022new, galis2022fast}, wash trading \cite{von2022nft}, and anonymity \cite{songzkdet, battah2020blockchain}. However, the study of the NFT-to-Asset connection that decides the value of NFTs, is still unaware, which hinders us from understanding whether NFTs' assets are stored safely in practice. Therefore, we conduct a systematic measurement and analysis of the NFT-to-Asset connection of NFTs on Ethereum. To be more specific, our study attempts to answer the following research questions:
    
    \begin{itemize}[label={\ding{228}}]
        \item [\textbf{RQ1}] Where are the NFTs' off-chain assets? Are NFTs still in peril of aggregation? 
        \item [\textbf{RQ2}] What is the degree that NFTs are connected to the off-chain assets? At which stage do NFTs lost connection with the assets?
        \item [\textbf{RQ3}] To what extent are NFT assets duplicate? At which step does the repetition occur? Which sort of NFTs are highly repetitive?
    \end{itemize}
    
    We initiate our study with non-trivial efforts by collecting NFT contracts based on \textsc{Etherscan}'s "\textit{NFT}" contract tag, and subsequently retrieve the data from on-chain \textit{TokenURI} to off-chain \textit{Metadata} and \textit{Asset}. We have to overcome several challenges to improve the measurement accuracy, \textit{e.g.}, IPFS instability and the changing availability of both IPFS and servers' data. In our measurement framework, we propose to leverage multiple gateways to enlarge the data coverage and extend a longer measurement period to cover as much of the time as possible  when servers went online (refer to \cref{sec:Factors}).  Our  measurement framework focuses on the angles of storage (refer to \cref{sec:thestorage}), accessibility (refer to \cref{sec:accessibility}) and the content (refer to \cref{sec:assets}), aiming to uncover the sustainability, the availability and redundancy of NFT. Through this multi-perspective profiling, we present a thorough and comprehensive understanding of the NFT-to-Asset connection, where the interesting and valuable findings reveal the present stage of NFTs in the wild is concerned. 
    
    The connection between NFTs and their assets is actually fragile, bringing at least but not limited to the following concerns: \ding{182} \textbf{Lost and inaccessible} (refer to \cref{sec:DoseNFTpreferdecentralizedstorage}, \cref{sec:2stagepath}, \cref{sec:Accessibilityateachstages}). The value of NFTs heavily depends on the safety of the underlying asset. Under the existing off-chain settings, owning NFTs doesn't really possess the assets. We find that assets of 3,118 (25.24\%) NFT contracts are not accessible. Whenever the actual asset is lost, ownership of such NFTs becomes worthless. Even with the more recognized decentralized storage, infrequently accessed data is vulnerable to loss or inaccessibility.  \ding{183} \textbf{Counterfeit} (refer to \cref{sec:duplicatedurl}, \cref{sec:identicalcontent}). As the transparent nature of the public blockchain, assets of NFTs can be accessed by anyone, which can be easily used to generate new NFTs by others. Such hypothesis is proved by our study. We find that 2,653 (21.48\%) NFT contracts include duplicated assets. \ding{184} \textbf{Unclear custodian responsibility} (refer to \cref{sec:clustercecntralized}). Once smart contracts are deployed on blockchain, NFTs' owners are not necessarily responsible for maintaining the assets by themselves. We observe that 4,801,699 (77.02\%) NFTs of centralized storage are aggregated on merely 1,184 third-party storage sites. The lost of these data renders unclear responsibility among buyers, sellers, and third-party storage platforms.

    In summary, our contributions are as follows:
    
    \begin{itemize}
        \item To the best of our knowledge, our work is the first to present an extensive characterization of the NFT-to-Asset connection fragility between on-chain NFTs and the associated off-chain assets from three perspectives: storage, accessibility and duplication.
        \item In order to overcome challenges of affecting the measurement accuracy, \textit{e.g.}, IPFS instability and the changing availability of both IPFS and servers' data, we propose to improve the measurement accuracy by leveraging multiple gateways to enlarge the data coverage and extend a longer measurement period with non-trivial efforts,.  
        
        \item Our study provides empirical evidence to suggest that current NFT-to-Asset connection is very fragile, which has non-trivial effect for NFT safety and trust. Key interesting and unexpected findings are listed  as follows:
        
            \begin{itemize}[leftmargin=0.5cm]
               
                \item \textbf{Off-chain assets are aggregated} (response to \textbf{RQ1}). We uncover evidence that 10 centralized platforms account for hosting 79.04\% of NFTs' off-chain assets, showing that the NFT off-chain assets are indeed centralized.
                 \item \textbf{Instability of decentralized storage} (response to \textbf{RQ1}). We show that decentralized storage has not yet achieved its goal of providing everlasting data storage in practice as we find that only 33.77\% of NFTs' decentralized assets can be fully retrieved. 
                \item \textbf{The 2-stage storage state are inconsistent} (response to \textbf{RQ2}). We identify a deceptive practice that though 48 NFT collections appear to be decentralized (according to their on-chain URLs), their underlying assets are actually centralized.
                \item \textbf{Duplicated NFTs} (response to \textbf{RQ3}). We find that NFT assets are severely duplicated. Of the NFT contracts that contain duplicated assets, 36.82\% of such have duplication rate greater than 95\% regarding to assets' URL.
               
            \end{itemize}

        \item Our observations shed light on the fragility along the NFT-to-Asset connection, which brings important implications to the NFT community: \ding{182} On the one hand, NFT holders or prospective investors need to pay closer attention to the risks associated with NFT-to-Asset connection and the uniqueness of their assets. \ding{183} On the other hand, NFT developers can enhance the strength of the NFT-to-Asset connection by integrating the NFT with other techniques, such as storage incentive, digital rights management, and binding mechanism, to increase the sustainability, traceability, and uniqueness of the NFT. \ding{184} And for the NFT applications, a more precise division of storage responsibilities and usage rights with users is needed to improve the NFT's credibility.
    \end{itemize}


\section{Background}
\label{Background}
    \subsection{Non-Fungible Token}
        \subsubsection{Token Standard} NFTs are driven by standard smart contracts that conform to specific interfaces, \textit{e.g.}, \texttt{ERC-721} \cite{ERC721}, \texttt{ERC-1155} \cite{ERC1155}, \texttt{ERC-998} \cite{ERC998}. Under the same token standard, NFTs have identical functions to achieve compatibility. \texttt{ERC-721} refers to the original \textit{Non-Fungible Token Standard} that each token is unique and not interchangeable. Descriptive information about NFTs is recorded in the \textit{metadata} field of NFT, including function  \textit{name()}, \textit{symbol()} and \textit{tokenURI()}. The return value of \textit{TokenURI} is usually an external URL pointed to a \texttt{JSON} table that is saved in another repository. Another small proportion of \textit{TokenURI} will return the metadata or the asset directly. Later when \texttt{ERC-721} became unable to meet the increasingly complex requirements of the blockchain application, the \textit{Multi Token Standard} \texttt{ERC-1155} is proposed to reduce the overhead of token transfer. \texttt{ERC-1155} is an extension of \texttt{ERC-721} that a group of fungible tokens and non-fungible tokens are deployed within one contract. The metadata of \texttt{ERC-1155} NFTs can be derived with function \textit{uri()} via an external URL. In addition to the ability to manage multiple tokens simultaneously, \textit{Composable Non-Fungible Token Standard} \texttt{ERC-998} additionally defines affiliation between NFTs, meaning that NFTs can be traced upstream and downstream. In other words, an NFT can be included in another NFT within \texttt{ERC-998} (\textit{e.g.}, \textsc{CryptoRome} \cite{CryptoRome}, \textsc{Mokens} \cite{mokens}, \textsc{Bitizens} \cite{bitizens}). Notice that \texttt{ERC-721} and \texttt{ERC-1155} provide an iterable function to obtain \textit{TokenURI}, while \texttt{ERC-998} does not. 
        
        Other protocols improve NFTs' compatibility and interoperability. For instance, the \textit{Semi-Fungible Token Standard} \texttt{ERC-3525} \cite{ERC3525} defines that each NFT can be split and reconstituted. The \textit{NFT Royalty Standard} \texttt{ERC-2981} \cite{ERC2981} defines the way to pay to a certain address with a certain value, which overcomes the obstacle that sale royalty is unable to be transferred across various platforms.

        \begin{figure*}[htbp] 
            \centering
            \includegraphics[width=0.95\textwidth]{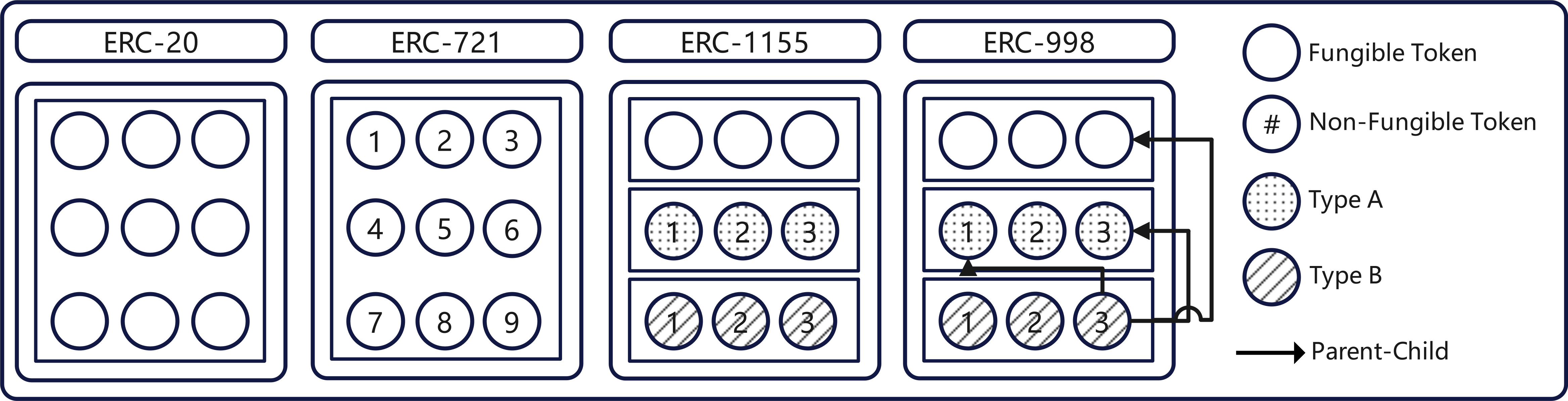}
            \caption{The NFT token standard.}
            \label{ERC}
        \end{figure*}
        
        \subsubsection{NFT-to-Asset Connection} As stated in the introduction, NFTs use a multi-hop URL to link to the asset. A typical NFT storage structure is that the \textit{TokenURI} is stored on the blockchain, which points to the \textit{Metadata} corresponding to the NFT. \textit{Metadata} is usually a \texttt{JSON} table with s certain schema, including title, type, image link, properties, etc. The \textit{Metadata} contains the \textit{AssetURI} to the final storage location of the underlying digital asset. The structure is illustrated in Figure~\ref{nft2asset}. Not all NFTs conform to this storage structure. For some NFT data that is stored on the blockchain, the \textit{TokenURI} will return the NFT data directly; Other NFTs may omit the \textit{TokenURI} and save \textit{Metadata} on the blockchain; Also there exists some NFTs that have no metadata but the \textit{TokenURI} is pointed to the assets. Saving data on-chain is undoubtedly the optimum solution, as data will be verified and permanently hosted by nodes included in the blockchain network.

    \subsection{Data Storage}
         \subsubsection{Centralized Storage} Centralized storage refers to the regular web hosting that the asset URL is on the server and the server is controlled by the server manager. These kinds of URLs are \textit{http://path} or \textit{https://path} style. These URLs are on the NFT marketplaces (\textit{e.g.}, \textsc{Opensea} \cite{opensea}, \textsc{Nifty Gateway} \cite{niftygateway}, \textsc{Rarible} \cite{Rarible}), NFT application developers, cloud services (\textit{e.g.}, \textsc{Google Cloud} \cite{cloudgoogle}, \textsc{Amazon Web Services} \cite{aws}). Hosting on such servers, the file size is almost unlimited and the data response becomes faster. However, since data is highly centralized and the NFT owner has no control over the data, once the servers stop responding or are attacked, the lovely images and vivid works of NFTs will disappear. The data also faces the risk of being tampered with, and there is no corresponding measure to ensure the rights of NFT users. For example, the NFT originator \textsc{CryptoKitties} \cite{cryptokitties} adopts centralized storage. Kitties on-chain is merely a series of numbers, and their corresponding identities and properties are on the company's servers, which is isolated from its on-chain notation.
            
        \subsubsection{Decentralized Storage} 
        \label{DecentralizedStorage}
        Unlike centralized storage, decentralized storage networks are free-used, trust-minimized, and tamper-resistant. Data are replicated among multiple nodes and the data will never disappear as long as there is a node in the world that stores this file. As summarized in Table~\ref{centalizedvsdecentralized}, compared with centralized storage that uses file path as URL, decentralized storage uses \textit{Content Identifier} (\texttt{CID}) calculated by the hash value of the file together with certain prefix information, so that the file cannot be tampered with. \textsc{Filecoin} and \textsc{Arweave} are both the incentive layer based on \textsc{IPFS}, aiding with blockchain technology, so as to motivate miners to actively keep data correctly and long-lastingly. Nodes of \textsc{IPFS} voluntarily store files that are broadcasting in the network and when the storage space is exhausted, the previous files are deleted and new files are stored. URLs of such storage are usually in the format as \textit{ipfs://<cid>} (\textsc{IPFS}), \textit{ar://<cid>} (\textsc{Arweave}), or the \texttt{CID} with or without gateway directly. 
        
        Although decentralized storage has made great progress in tamper resistance and data persistence, there are still some risks in this method. \ding{182} \emph{The infrequently requested data will also lost as the earlier file will be deleted when it achieved the upper limit of nodes' storage capacity}. For this reason, there are now \textsc{IPFS} storage custodial services like \textsc{Pinata} \cite{pinata}, which help \textsc{IPFS} data to be hosted forever and reduce the probability of loss. \ding{183} Even if the data exists in \textsc{IPFS}, \emph{it can also be massively aggregated in the \textsc{IPFS} nodes of some NFT applications}. Since these NFTs are often obtained directly through their gateways, these data will be rarely backed up in other \textsc{IPFS} nodes, increasing the risk of loss. \ding{184} \emph{The response time of \textsc{IPFS} can be extremely laggy and the possibility of successful retrieval is not guaranteed} (a brief demonstration is given in \cref{IPFSInstability}) since the node does not necessarily store the requested file, nor does it necessarily be online.
        
        \begin{table*}[htbp]
        \resizebox{\textwidth}{!}{
        \begin{tabular}{@{}cccccc@{}}
        \toprule[1.5pt]
        \textbf{Storage}   & \textbf{Identifier} & \textbf{Address by}  & \textbf{Modifiable}   & \textbf{Permanent} &  \textbf{Example}                                               \\ \midrule
        \multirow{2}{*}{Centralized}   & \multirow{2}{*}{\begin{tabular}[c]{@{}c@{}}URL\\ (File Path) \end{tabular}} & \multirow{2}{*}{Location} & \multirow{2}{*}{\Checkmark} & \multirow{2}{*}{\XSolid} & \textit{\textbf{https://}i.imgur.com/amSZEPe.jpg}  \\ \cmidrule(l){6-6} &  &  &  &  & \textit{\textbf{https://}cryptoz.cards/data/0} \\ \midrule
        \multirow{2}{*}{Decentralized} & \multirow{2}{*}{\begin{tabular}[c]{@{}c@{}}CID\\ (File Hash)\end{tabular}} & \multirow{2}{*}{Content}  & \multirow{2}{*}{\XSolid}  & \multirow{2}{*}{\XSolid} & \textit{\textbf{ipfs://}QmNRtxvce6G1pnAs7HLoS152XeF1fr7EGXTQq9aKXJUMov} \\ \cmidrule(l){6-6} & & & & & \textit{\textbf{ar://}7cJVjjwuVa-GJhzS6w9GllPl0iqnXVuJciV4NVS7xp4}     \\ \bottomrule[1.5pt]
        \end{tabular}}
        \caption{Centralized storage \textit{vs}. Decentralized storage.}
        \label{centalizedvsdecentralized}
        \end{table*}
        
    \subsection{Asset Consistency}
        Asset Consistency refers to that the data can be verified as complete and reliable, which provides additional assurance. There are roughly two sorts of approaches to achieve data consistency: to save the data on-chain or save data hash on-chain.
        
        \subsubsection{Hash value} Recording the cryptographic hash of the data on the blockchain is the most widely adopted approach to preserving data integrity. \texttt{CID} of decentralized storage itself is the hash of the data. Once data is published, it cannot be modified. Data can be updated only by publishing another modified data, and the \texttt{CID} becomes different. Hence, the \textit{InterPlanetary Name System} (\textsc{IPNS}) of \textsc{IPFS} is used to manage the version update \cite{IPNS}. But even if the hash value itself is much smaller than the original data, the cost of storing hashes is still expensive. Suppose that the NFT data is hashing via \texttt{SHA-256} and produces a 32-byte hash; the number of NFTs published in this project is 1,000; the current gas fee is around 150 gwei \cite{EtherscanGasTracker}. Besides, the Ethereum yellow book stated that 20,000 units are paid for \textit{SSTORE} operation per 32 bytes \cite{EthereumYellowPaper}. The original storage cost of these NFTs is thereby $1,000\times20,000\times150 / 10^9 = 3$ ETH, which is \$79,815. Consequently, a modified method is to store the data hash in groups, such as using the Merkle root hash of a batch of data. For example, the digital art collection \textsc{Hashmasks} \cite{HashmasksProvenanceRecord} generates hash provenance by hashing the concatenating images, and the image's URL is not saved on the blockchain.
        
        \subsubsection{On-chain Storage} To be more straightforward, data on the blockchain live permanently and will not be tampered with, which is the optimal solution for NFT data storage. For example, \textsc{Akomba Commemorative Token} (\texttt{AkCT}) \cite{Akomba} are quotes from crypto-celebrities; \textsc{ColorverseFounder} (\texttt{CVF}) \cite{colorverse} and \textsc{Mandala Tokens} (\texttt{MANDALA}) \cite{Mandala} are the \texttt{SVG}s of the mosaic of color blocks. A more applicable and gas-saving solution is to deploy the \texttt{SVG} generators on-chain, where the image can be derived with the NFT properties directly, such as the top decentralized exchange \textsc{Uniswap V3} (\texttt{UNI-V3-POS}) \cite{UniswapSVGNFT} and DeFi-NFT \textsc{Aavegotchi} (\texttt{GOTCHI}) \cite{AavegotchiOnchainSVGs}.

    
 \section{Dataset and Experimental Setup}
    \subsection{Data Collection}
    \label{sec:assetcollection}
        According to the NFT-to-Asset structure, data collection is divided into 4 sub-processes:
        
            \vspace{1mm}
            \noindent{\textbf{NFT Contract.}} In the beginning, we obtain 12,353 NFT contract addresses from \textsc{Etherscan}, where NFT contracts are tagged out \cite{NFTList}. NFTs in a same contract are also called an NFT \textit{collection}.
            
            \vspace{1mm}
            \noindent{\textbf{The \textit{TokenURI}.}}  The \textit{TokenURI} collection has two data sources: one from the \textsc{Ethereum Mainnet} and one from \textsc{Opensea}. We make our endeavor to ensure that the data comes from the \textsc{Ethereum Mainnet} for the most part, as the \textit{TokenURI} provided by other platforms isn't always consistent with the one in Ethereum. For example, the \textsc{Waifus} NFTs have no publicly available \textit{TokenURI} according to its contract, but it is given by \textsc{OpenSea} that the \textit{TokenURI} is \textit{https://api.waifusion.sexy/v1/opensea/i}, which is somewhat in conflict with the record on the blockchain. In order to retrieve \textit{TokenURI} from Ethereum, we build a full archive \textsc{OpenEthereum} node to join the Ethereum mainnet for synchronization, which enables full contract state information\cite{OpenEthereum}. The workflow starts with requesting the total supply of NFTs in an NFT contract (function in \texttt{ABI}: \textit{totalSupply()}). If the contract is not of empty supply, then we sequentially request the id of NFT and then use the token id to retrieve \textit{TokenURI} and its ownership(function in \texttt{ABI}: \textit{tokenByIndex()}, \textit{tokenURI()}, \textit{ownerOf()}).
            
            However, a portion of \textit{TokenURI}s are unable to be obtain from Ethereum for the following reasons: \ding{182} the NFT contract does not always conform to the standard form, i.e. the \texttt{ERC-721}, \texttt{ERC-1155}, and some variants; \ding{183} contracts are not open-source, which is not available on \textsc{Etherscan}; \ding{184} the \textit{TokenURI} related function can not be called from outside the contract, which prevents us from invoking the contract function to collect the information. Therefore, for contracts that are not accessible from Ethereum, we obtain it with \textsc{OpenSea} "\textit{asset}" API \cite{OpenSeaAssetAPI}, and extract the \textit{TokenURI} from \textit{token\_metadata} field. But unfortunately, \textsc{OpenSea}'s API has a maximum access limit, which is controlled by parameter "\textit{offset}" and "\textit{limit}". That is, only the first 10,050 NFTs can be obtained. 23 contracts exceed the upper limit of \textsc{OpenSea} during this process, thus we try to collect from these contracts manually. \textit{TokenURI} of 10 contracts are successfully retrieved, 6 of them can be obtained but the value is empty, hence the full NFT dataset of these contracts is obtained. Besides, 7 of them are unable to get due to the non-public contracts' codes.
            
            
            \vspace{1mm}
            \noindent{\textbf{The \textit{Metadata}.}} Metadata is obtained by requesting the \textit{TokenURI}. For location-based URL (in HTTP style), we request the data ordinarily. For content-based URL (\textsc{IPFS} and \textsc{Arweave}'s \texttt{CID}), we request the data from hybrid sources: a locally build \textsc{IPFS} node, and the public gateways (\texttt{ipfs.io} \cite{ipfsiogateway}, \texttt{gateway.pinata.cloud} \cite{pinatagateway}, \texttt{infura-ipfs.io} \cite{infuragateway} and \texttt{dweb.link} \cite{dwebgateway}), considering the response instability of \textit{IPFS} (refer to \cref{IPFSInstability}).
            
            \vspace{1mm}
            \noindent{\textbf{The \textit{Asset}.}}
             A standard \texttt{ERC-721} metadata format contains 2 URLs, one is in the \textit{image} field and the other is in the \textit{external\_url} field \cite{ERC721}. Both the URLs are usually in different locations, such as an \textit{image} of the IPFS-style URL and an \textit{external\_url} of the location-based URL for backup. We thus extract the \textit{AssetURI} automatically from the following fields: \textit{image}, \textit{external\_url}, \textit{url}, and \textit{animation\_url}. Then we manually collect URLs from the irregular metadata. The following process is the same as that in the \textit{Metadata} part, which is to collect content-based URLs and location-based URLs separately to obtain the asset.

    \subsection{Factors Affecting Measurement Accuracy}
    \label{sec:Factors}
        \subsubsection{\textbf{\textsc{IPFS} Instability}}
        \label{IPFSInstability}
            As a large amount of NFTs' data is hosted on \textsc{IPFS}, the accuracy of the measurement results highly depend on the availability of \textsc{IPFS} data. During data collection, we observe that the data availability of \textsc{IPFS} exists inconstant characteristics (refer to \cref{DecentralizedStorage}), which is reflected in the following aspects: \ding{182} There are plenty of cases where data can't be found. \ding{183} The accessibility of the same data at the same time with different public gateways as well as the \textsc{IPFS} nodes built in our experiment shows different results. \ding{184} The data request via \textsc{IPFS} is extremely slow, rendering additional response failure. One reason is that no more nodes have saved this data and the data is indeed lost forever. Another reason is that the data is not lost, but possibly nobody online has it, or the node with the data is hidden behind the internal network, or the node with data does not publish the data in the way that an ordinary node can find it.
            
\begin{figure*}[htbp]
                \centering
                    \includegraphics[width=0.75\textwidth]{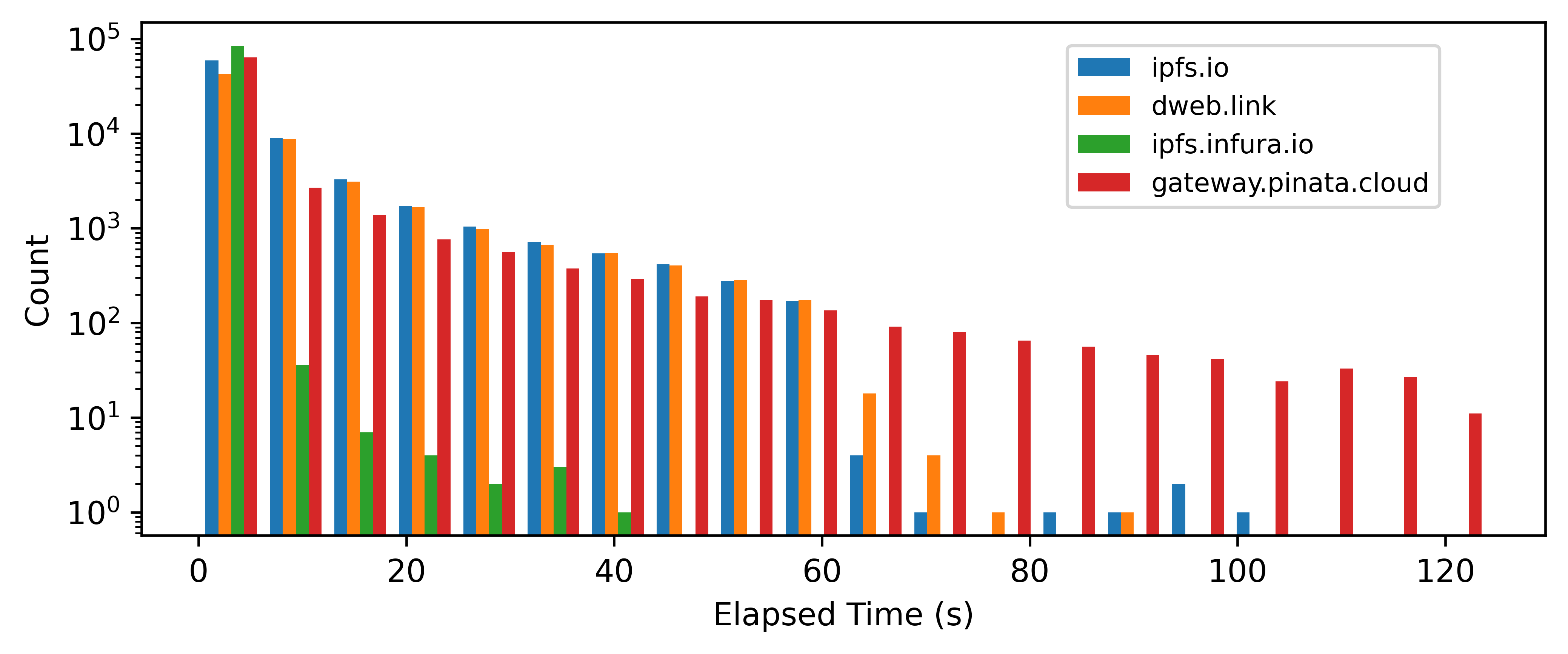}
                \caption{IPFS response elapsed with different gateways.}
                \label{ipfselapse}
            \end{figure*}

            \textbf{Solution:} The \textsc{IPFS} data is collected with multiple gateways to enlarge the data coverage. The response duration and the hit rate of different \textsc{IPFS} gateways is presented in Figure~\ref{ipfselapse} and Table~\ref{IPFSresult} to describe the above instability. It can be seen from Figure~\ref{ipfselapse} that the response time of different gateways varies greatly. The fastest response is \textsc{Infura}, \textsc{IPFS.io}, and \textsc{Dweb.link} almost cut off after a minute, while \textsc{Pinata} has the heaviest tail. The numerical results are summarized in Table~\ref{IPFSresult}. Although the time difference is large, the hit rates of the gateways differ only 3.43\% at most, which is not higher enough but not significantly different. Overall, 88.18\% of the URLs are fully obtained by all gateways, and the union of URLs that are successfully obtained by each gateway covers 97.98\% of the sample set. Therefore, the above data shows that our method is effective to overcome the above instability.
            
            \begin{table}[htbp]
                \resizebox{14cm}{!}{
                    \begin{tabular}{lrrrrrllr}
                    \toprule[1.5pt]
                    \multicolumn{1}{c}{\multirow{2}{*}{\begin{tabular}[c]{@{}c@{}} \textbf{Gateways}\\ \textit{https://\textless{}gateway\textgreater{}/ipfs/\textless{}cid\textgreater{}}\end{tabular}}} & \multirow{2}{*}{\begin{tabular}[c]{@{}c@{}} \textbf{Average} \\ \textbf{Response}\end{tabular}} & \multirow{2}{*}{\textbf{Hit Rate}} & \multicolumn{3}{c}{\textbf{Failure Type}}  &  & \textbf{Hits} & \textbf{\# of URL} \\ \cmidrule{4-6} \cmidrule{8-9} 
                    \multicolumn{1}{c}{}   &   &   & \multicolumn{1}{c}{\textbf{Server}}  & \multicolumn{1}{c}{\textbf{Client}}  & \multicolumn{1}{c}{\textbf{Other}} &  & 4    & 88.18\%   \\ \cmidrule{1-6}
                    \texttt{ipfs.io}               & 5.19s   & 93.41\%   & 4,159 (63.07\%) & 2,387 (36.20\%) & 48 (0.07\%)  &  & 3    & 5.89\%    \\
                    \texttt{dweb.link}             & 6.32s   & 93.15\%   & 4,848 (70.80\%) & 1,943 (28.38\%) & 56 (0.08\%)  &  & 2    & 1.93\%    \\
                    \texttt{ipfs.infura.io}        & 1.45s   & 92.60\%   & 4,661 (63.02\%) & 2,691 (36.38\%) & 44 (0.06\%) &  & 1    & 1.79\%    \\
                    \texttt{gateway.pinata.cloud}  & 3.32s   & 96.03\%   & 1,417 (35.65\%) & 2,427 (61.05\%) & 131 (3.30\%) &  & 0    & 2.27\%    \\ \cmidrule{1-6} \cmidrule{8-9}
                    \textbf{Aggregate Hit Rate} & \multicolumn{2}{r}{97.98\%} \\ \bottomrule[1.5pt]
                    \end{tabular}}
                \caption{Comparison of \textsc{IPFS} gateways. We sample 100,000 IPFS-style URLs evenly from a set including both \textit{TokenURI} and \textit{AssetURI}, and 4 active gateways from the \textsc{IPFS} gateway selector \cite{gatewaychecker}. The \textit{CID} of IPFS is extracted from the sampled URLs, and then combine with the selected gateways to form a new \textsc{IPFS} URL for the serial request. The process is performed simultaneously among gateways to eliminate the network effect caused by the client.}
                \label{IPFSresult}
            \end{table}

        \subsubsection{\textbf{The Changing Availability}}
        We note that both \textsc{IPFS} and servers' data availability fluctuates at different moments. For example, NFTs from \textsc{Unlock Protocol} \cite{unlockprotocol} fail to be accessed until the last round of data collection. The low availability occurs randomly and is hard to define as data loss as it will cause data loss to be overestimated.
        
        \textbf{Solution:} We extend the data collection process into a six-month period to cover as much of the time as possible when servers went online.  In each collection cycle, we first remove invalid files by keywords identification, such as "\textit{Gateway Time-out}", "\textit{Server Error}", \textit{etc.}, and then keep requesting the uncovered data. New NFTs are acquired after each cycle of data collection, reducing false-positive judgement of data loss.

    \subsection{Overall Statistics}
    \label{sec:overallstatistics}

        We use the results of calling the \textit{totalSupply} function and the \textit{TokenURI} function of NFT contract to describe the basic situation of the NFT contract and the proportion of data collected from different sources. The contracts are divided into 5 circumstances. 
        \ding{182} \textit{Complete}: the \textit{totalSupply} is identical to the available \textit{TokenURI}; \ding{183} \textit{Not Callable totalSupply}: The \textit{totalSupply} is not callable or the return is null; \ding{184} \textit{Empty Supply}: the \textit{totalSupply} is zero, indicating that the contract is empty; \ding{185} \textit{Great Supply}: The \textit{totalSupply} is greater than 1 million; \ding{186} \textit{Not Callable TokenURI}: the \textit{totalSupply} is a non-zero value, but \textit{TokenURI} is not callable or the returns are null.  
        
        According to Table~\ref{Contract}, from the results of the initial call to the \textsc{Ethereum Mainnet}, there are 8,056 contracts in \textit{Complete} status of which we successfully collect the entire \textit{TokenURI} information. There are 752 contracts in \textit{Not Callable totalSupply} status, and there are 685 \textit{Not Callable tokenURI} contracts. There are two reasons why the functions are not callable, one is that the contract is not in the standard format, another is that it might be the \texttt{ERC-998} NFT contract as it does not provide an interface of \textit{totalSupply()} or \textit{tokenURI()}. Besides, the number of \textit{Empty Supply} and \textit{Great Supply} contracts are 2,851 and 9 respectively. We thereby request the \textit{Not Callable totalSupply} and \textit{Not Callable TokenURI} contracts to \textsc{Opensea}. Among the 752 \textit{Not Callable totalSupply} contracts, 549 of which then become \textit{Complete} and 203 of which are \textit{Empty Supply}; Among the 685 \textit{Not Callable tokenURI} contracts, 615 of which then become \textit{Complete} and 70 of which are \textit{Empty Supply}. 
        
        In summary, we have 6,234,141 NFTs according to 12,353 contracts. 10,916 (88.37\%) contracts information are collected via \textsc{Ethereum Mainnet} directly and 1,437 (11.63\%) contracts information are collected from \textsc{Opensea}. We find that 3,123 (25.28\%) contracts are empty, \textit{i.e.}, a zero \textit{totalSupply}, and hence they will not be measured in the following. Table~\ref{NFTdata} provides the data collection from \textit{TokenURI} to the \textit{Asset} in general. Ideally, each step of the NFT-to-Asset path should be injective, that is the number of \textit{TokenURI}s, \textit{Metadata}, \textit{AssetURI}s and \textit{Asset}s should be equal, ignoring the very few NFTs which are exceptional to this normal path. However, we find that \emph{the decrease of the un-collected data in the next stages is greater than the number of those exceptional NFTs. The number of assets collected at the end of the path is only 45.75\% of that of NFT, indicating that the issue of ineffective URL is more severe than expected}. A detailed analysis of NFT data loss is presented in \cref{sec:accessibility}.

        \begin{table}[htbp]
        \parbox[b]{0.47\linewidth}{
        \centering
        \resizebox{7cm}{!}{
        \begin{tabular}{@{}lrrrrrr@{}}
        \toprule[1.5pt]
        & \ding{182} & \ding{183} & \ding{184} & \ding{185} & \ding{186} & \textbf{Sum}    \\ \midrule
        \textbf{Ethereum Mainnet} & 8,056  & \textbf{752} & 2,851 & 9  & \textbf{685}  & 12,353 \\ \midrule
        \textbf{Opensea} + \ding{183} & 549 & \textbackslash{} & 203 & \textbackslash{} & \textbackslash{} & \textbf{752}    \\
        \textbf{Opensea} + \ding{186} & 615 & \textbackslash{} & 70  & \textbackslash{} & \textbackslash{} & \textbf{685}    \\ \midrule
        \textbf{Aggregate Reseult}  & 9,195  & \textbackslash{}  & 3,123 & 9 & \textbackslash{} & 12,353 \\ \bottomrule[1.5pt]
        \end{tabular}}
        \caption{Contract information and data source. \ding{182} \textit{Complete}; \ding{183} \textit{Not Callable totalSupply}, \ding{184} \textit{Empty Supply}; \ding{185} \textit{Great Supply}, \ding{186} \textit{Not Callable TokenURI}.}
        \label{Contract}
        }
        \hfill
        \parbox[b]{0.47\linewidth}{
        \centering
        \resizebox{6cm}{!}{
        \begin{tabular}{@{}lrrr@{}}
        \toprule[1.5pt]
                 & \textbf{Row Count} & \textbf{Data Size}  & \textbf{Hit Rate} \\ \midrule
        \textbf{Contract} & 12,353    & \textbackslash{} & \textbackslash{} \\
        \textbf{NFT}      & 6,234,141 & \textbackslash{} & \textbackslash{} \\
        \textbf{\textit{TokenURL}} & 5,846,287 & 1.53 GB          & 93.78\%         \\
        \textbf{\textit{Metadata}} & 4,566,766 & 119.88 GB        & 78.11\%   \\
        \textbf{\textit{AssetURL}}   & 3,992,859 & 413.54 MB        & 87.43\%   \\
        \textbf{\textit{Asset}}    & 2,851,894 & 3.29 TB          & 71.42\%   \\ \bottomrule[1.5pt]
        \end{tabular}}
        \caption{Summary of collected NFTs and the associated data.}
        \label{NFTdata}
        }
        \end{table}

\section{The Storage}
\label{sec:thestorage}
    In this section, we make an assessment of the overall storage situation of the NFT ecosystem and study the impact the vulnerable storage that has on the NFT. To this end, we first investigate the proportions of NFTs storage on centralized servers and decentralized storage networks. We then conduct a progressively deeper investigation towards the exact server clusters where NFTs' associated data is hosting on, aiming to identify the severity of data aggregation. Next, we take the 2-stage storage as a whole to investigate the stability and consistency of the NFT-to-Asset connection.

    \subsection{The 2-stage storage.}
        To ascertain whether the NFT storage is sustainable, we first look at the location distribution in the two storage phases and provide the correlation between the two storage layers. Then, we conduct a detailed study of the scenario in which NFTs for centralized storage are constantly clustered on a few number of large platforms.
        
        \subsubsection{Do NFTs prefer decentralized storage?}
        \label{sec:DoseNFTpreferdecentralizedstorage}
\begin{figure}[htbp]
            \centering
            \includegraphics[width=13cm]{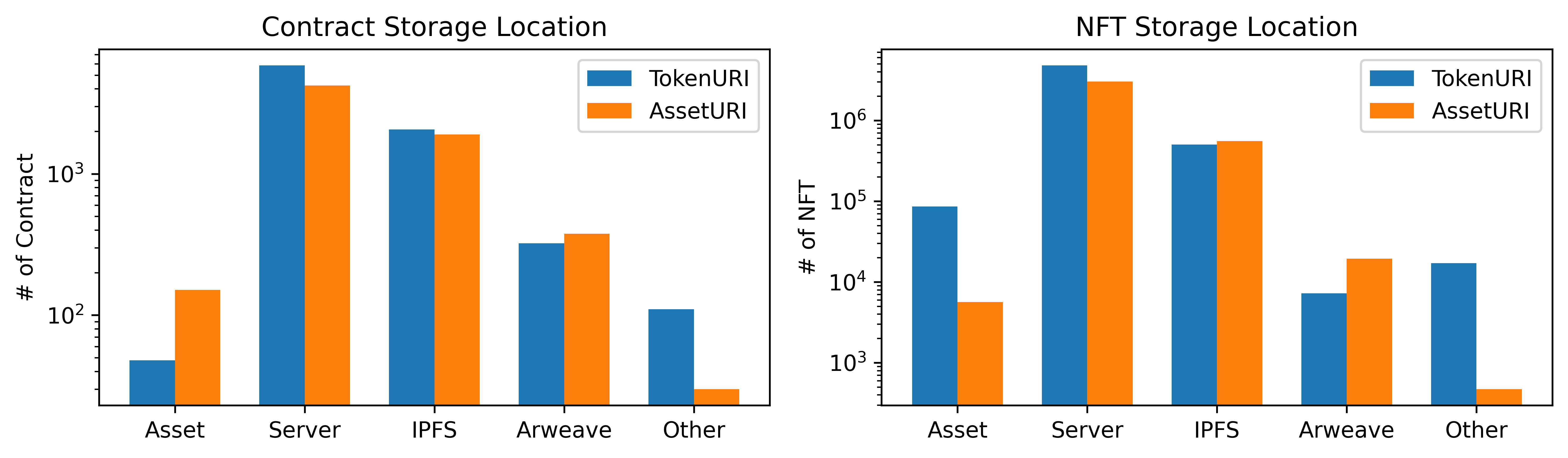}
            \caption{The 2-stage storage distribution by the number of contracts and NFTs respectively.}
            \label{storagedistribution}
        \end{figure}

        Figure~\ref{storagedistribution} shows the number of contracts and the number of NFTs stored at different locations. The number of contracts refers to the left axis, and the number of NFTs, which is scaled after the base-10 exponential transform, refers to the right axis. The blue bar represents the \textit{TokenURI} at 1st-stage, and the orange bar represents the \textit{AssetURI}. 
        The \textit{Asset} of 1st-stage storage indicates that the data of NFTs is directly saved on-chain, and of 2nd-stage refers that the NFT's metadata is exactly the NFT data. The \textit{Other} includes some niche forms of NFT data, such as a local path of a computer, a string of numbers, peoples' names, dates, unknown coding, etc. If the NFT is labeled as \textit{Empty}, its return value of function \textit{tokenURI()} on the blockchain is empty. 
        
        According to Figure~\ref{storagedistribution}, we find that centralized storage is still the primary choice for NFTs today, both in terms of the number of contracts and the number of NFTs. In the 1st-stage of storage, a total of 5,862 contracts and 4,802,766 NFTs of data are stored on centralized servers. Decentralized storage is mainly undertaken by IPFS and \textsc{Arweave} networks. In this phase, a total of 2,385 contracts and 510,154 NFTs use decentralized storage. Therefore, the number of contracts for decentralized storage is only 40.69\% of that for centralized storage, and the number of NFTs in the contract for decentralized storage is only 10.62\% of that for centralized storage, indicating that the average number of NFTs in the contract of decentralized storage is lower. Among decentralized storage, \textsc{IPFS} has a much higher adoption rate than \textsc{Arweave}, where \textsc{Arweave} has only 1.43\% of \textsc{IPFS}'s NFT. In addition, we find that one contract selected the only other decentralized storage platform \textsc{Storj} \cite{storj} for storing NFT assets, while its metadata is stored on the server of \textsc{Mintable} \cite{mintable}, an NFT trading market.
        
        In the second phase of storage, most of the \textit{AssetURI}s are extracted from the \textit{Metadata}, that is the asset may have more than one URL. Assets are stored in different locations for backup to reduce the risk of asset loss. The storage of servers, \textsc{IPFS}, and \textsc{Arweave} in this phase is generally similar to that in the previous phase. The number of contracts and NFTs for centralized storage decreased obviously by 28.14\% and 37.21\% respectively compared with the previous stage, due to the interruption of some NFTs' URL paths or the unavailability of URLs. On the contrary, although the number of contracts for decentralized storage decreased by 4.65\%, the number of NFTs increased by 12.33\%. The number of \textsc{IPFS} contracts decreased to 1,898, but the number of NFTs increased to 553,601; \textsc{Arweave} has seen more significant growth, with both contracts and NFTs increasing to 376 and 19,469. However, the absolute value of growth is not as large as \textsc{IPFS} due to its smaller base. This indicates that both \textsc{Arweave} and \textsc{IPFS} are used as backup networks by a portion of the NFT of centralized storage, where \textsc{IPFS} is free and \textsc{Arweave} is paid at one time.
        
         In \textit{Other} category, the information returned by URLs is unable to trace towards the storage type in the next hop, such as the local address, meaningless repeated characters, string of digital coding, string of numbers, person's names and short sentences. It can be regarded as on-chain storage to a certain extent as the information is completely stored on the blockchain without having an off-chain URL. There are 17,096 NFTs (110 contracts) in \textit{Other} category in the \textit{TokenURI} storage. This number drops precipitously to just 472 \textit{Other} NFTs (30 contracts) in the \textit{AssetURI} storage, of which they are mainly the local address.
        
        \subsubsection{Connection of 2-stage storage} 
        \label{sec:2stagepath}
        We examine the relationship between the two stages of NFT storage, which is the direction of the 2nd-stage storage following specific 1st-stage storage, in order to determine the stability and consistency of the two-stage storage. We narrow the measurement scope to NFTs that successfully connect to the underlying assets, meaning that both \textit{TokenURI} and \textit{AssetURI} exist and the assets are able to be retrieved. NFTs with on-chain assets and NFTs that fail to connect to their assets are ignored since these assets have either been permanently stored in the blockchain or have been lost. Within a path, \textit{TokenURI} and \textit{AssetURI} may exist one or more different URLs to guarantee asset safety. We divide the of NFT-to-Asset connection into three categories:
        
        \begin{enumerate}
            \item \textit{Decentralized.} Both stages have at least one URL for decentralized storage. 
            \item \textit{Semi-decentralized.} One of the two storage contains no URL towards decentralized storage, and the other storage does: \ding{182} One case is that the 1st storage is decentralized and the 2nd-storage is centralized; \ding{183} the reverse of the previous one. Both cases weaken the relevance between NFTs and assets since the metadata is at risk of unauthorized modifications or be easily altered to another URL, since this change will not be recorded via blockchain consensus.
            \item \textit{Centralized.} Fully centralized NFTs contain no URL of a decentralized platform, giving them the weakest ability in tamper resistance. 
        \end{enumerate}
        
        
        The result is presented in Table~\ref{paths1}, with taking multi-URLs of \textit{AssetURI} into account. Among the 6,619 NFT contracts that are included in the measurement scope, there are 2,034 (30.73\%), 444 (6.71\%), and 4,141 (62.56\%) contracts in the \textit{Decentralized}, \textit{Semi-decentralized}, and \textit{Centralized} categories respectively. 68.59\% of the NFT collections use more than one platform for asset back-up, and the major portion of NFTs without assets back-up are stored on the server. \emph{However, we discover that there exists a deceptive practice, where 48 NFT collections that have decentralized metadata but centralized assets. In this way, the safety of the assets is not assured even though the metadata of NFTs is permanently preserved and tamper-resistant}. 
        
        \begin{table*}[htbp]
        \resizebox{\textwidth}{!}{
        \begin{tabular}{@{}llrllrllrllrllr@{}}
        \toprule[1.5pt]
        \multicolumn{6}{c}{\textbf{Decentralized}} & \multicolumn{6}{c}{\textbf{Semi-decentralized}} & \multicolumn{3}{c}{\textbf{Centralized}}   \\ \cmidrule(r){1-6} \cmidrule(rl){7-12} \cmidrule(l){13-15}
        \textbf{\textit{Metadata}} & \textbf{\textit{Asset}} & \textbf{\#}  & \textbf{\textit{Metadata}}   & \textbf{\textit{Asset}}   & \textbf{\#}  & \textbf{\textit{Metadata}}   & \textbf{\textit{Asset}}    & \textbf{\#} & \textbf{\textit{Metadata}}  & \textbf{\textit{Asset}}  & \textbf{\#}  & \textbf{\textit{Metadata}}  & \textbf{\textit{Asset}}  & \textbf{\#}  \\ \cmidrule(r){1-3} \cmidrule(rl){4-6} \cmidrule(rl){7-9} \cmidrule(l){10-12} \cmidrule(l){13-15}
        \texttt{AR}   & \texttt{AR}   & 48  & \texttt{IPFS} & \texttt{AR}  & 1 & \texttt{IPFS} & \texttt{SE} & \textbf{20} & \texttt{SE} & \texttt{AS}+\texttt{SE} & 1 & \texttt{SE}   & \texttt{SE}  & 1,851   \\
        \texttt{AR}   & \texttt{AS} & 1  & \texttt{IPFS} & \texttt{AS}  & 40 & \texttt{IPFS} & \texttt{SE}+\texttt{SE} & \textbf{24} & \texttt{SE} & IPFS+\texttt{SE}  & 95  & \texttt{SE}  & \texttt{SE}+\texttt{SE} & 2,284   \\
        \texttt{AR}   & \texttt{AR}+\texttt{AR}  & 1  & \texttt{IPFS} & \texttt{IPFS}  & 469  & \texttt{OC}  & \texttt{SE} & \textbf{4} & \texttt{SE} & \texttt{IPFS}+\texttt{SE}+\texttt{SE}  & 2   & \texttt{SE}    & \texttt{SE}+\texttt{SE}+\texttt{SE}  & 6  \\
        \texttt{AR}   & \texttt{AR}+\texttt{SE}  & 271  & \texttt{IPFS} & \texttt{IPFS}+\texttt{AS}  & 3 & \texttt{SE} & \texttt{AR} & 3  & & & & & & \\
        \texttt{OC}   & \texttt{IPFS} & 4 & \texttt{IPFS} & \texttt{IPFS}+\texttt{IPFS}  & 14 & \texttt{SE} & \texttt{AS} & 108 & & & & & & \\
        \texttt{OC}   & \texttt{AR}+\texttt{AR} & 3  & \texttt{IPFS} & \texttt{AR}+\texttt{SE} & 3 & \texttt{SE} & \texttt{IPFS} & 134 & & & & & & \\
        \texttt{OC}   & \texttt{AR}+\texttt{IPFS} & 13 & \texttt{IPFS} & \texttt{IPFS}+\texttt{SE} & 1,159 & \texttt{SE} & \texttt{IPFS}+\texttt{IPFS} & 20  &     & & & & & \\
        \texttt{OC}   & \texttt{IPFS}+\texttt{SE} & 1 & \texttt{IPFS} & \texttt{IPFS}+\texttt{SE}+\texttt{SE} & 3  & \texttt{SE} & \texttt{AR}+\texttt{SE} & 33 & & & & & & \\ \cmidrule(r){1-6} \cmidrule(rl){7-12} \cmidrule(l){13-15}
        \multicolumn{3}{l}{\textbf{Sum}} & \multicolumn{3}{r}{2,034} & \multicolumn{3}{l}{\textbf{Sum}} & \multicolumn{3}{r}{444} & \multicolumn{2}{l}{\textbf{Sum}} & 4,141 \\ \bottomrule[1.5pt]
        \end{tabular}}
        \caption{Distribution of NFT-to-Asset paths. The storage types are in line with \cref{sec:DoseNFTpreferdecentralizedstorage}: Server(\texttt{SE}), \textsc{Arweave}(\texttt{AR}), \textsc{IPFS}(\texttt{IPFS}), Asset(\texttt{AS}), On-chain(\texttt{OC}).}
        \label{paths1}
        \end{table*}

    \subsection{NFTs are clustering on centralized servers.}
    \label{sec:clustercecntralized}
        For NFT applications with demands for centralized storage, some of them will not maintain dedicated servers. Instead, a large proportion of NFTs choose to keep their data on third-party storage platforms, mainly in the following types: \ding{182} NFT marketplaces, which allow NFT developers and independent collectors to publish NFT works on the blockchain through the exchange market's own channels, while providing relevant data storage services; \ding{183} The cloud service providers, where users first publish their NFT works on to it and then upload the given URLs to the blockchain; \ding{183} The server of the application provider, where the data is maintained and managed by the application itself. \emph{In this way, NFT off-chain content is further clustered on a few servers, which goes against the primal decentralized intention of blockchain}. 
        
        Therefore, we intend to assess the extent to which NFTs are concentrated on a few servers. We examine the NFTs' hosting location by the domain from HTTP-based \textit{TokenURI}, and the domains with the highest concentration of NFTs are presented in Table~\ref{netloc}. Since the amount of NFTs in the contract varies greatly, the results are counted by contracts and NFTs separately. Of 4,801,699 NFTs (5,858 contracts) that adopt centralized storage, there are 1,184 different storage sites. \emph{The top-10 domains hold 79.04\% (65.67\%) of the total number of NFTs (contracts)}. Remarkably, the NFT marketplaces are oligarchical with \textsc{Opensea} \cite{opensea}, \textsc{NiftyGateway} \cite{niftygateway}, and \textsc{Mintable} \cite{mintable} considering the number of collections (in proportions of 17.94\%, 16.38\%, 15.47\%), and consequently have a greater influence on the safety of NFT ecosystem. While further examining the NFT storage on these marketplaces, we observe that the \textsc{Opensea} NFTs' metadata is on its own server, and the assets are hosted on \textsc{GoogleUserContent} (97.87\%) \cite{cloudgoogle} and \textsc{Opensea} (2.13\%). Nevertheless, the NFTs' metadata of 158 (10.28\%) contracts on \textsc{Opensea} fail to retrieve. The \textsc{NiftyGateway} target assets are saved on \textsc{Cloudinary} (99.47\%) \cite{cloudinary} and \textsc{AmazonAWS} (0.53\%) \cite{aws}. Still, we observe that the asset acquisition result accordingly of 25 (2.60\%) \textsc{NiftyGateway} contracts (1795 NFTs) is \textit{not found}. Compared to the above marketplaces, \textsc{Mintable}'s asset layer has no uniform schemes, and the NFTs' assets are directly hosted on the metadata-level URL's location in a one-hop way, or are on various storage such as \textsc{IPFS}, \textsc{Arweave}, \textsc{CloudFront} \cite{cloudfront}, \textsc{Imgur} \cite{imgur}, \textsc{AmazonAWS}, etc. The NFTs in 116 (12.8\%) contracts are not available, which is the highest loss rate among the three marketplaces.
        
        The aggregation of NFTs on servers is slightly scattered than that of contract, which is reflected in the smaller aggregation of the NFTs of the top-3 domains (in proportions of 12.77\%, 10.34\%, 10.27\%) and the larger aggregation of the other domains. However, the blockchain game \textsc{Gods Unchained} \cite{godsunchained} ranked 1st with 613,328 NFTs, though operating normally, \textit{TokenURI} recorded on Ethereum cannot access the data in the next level without exception. Another highly ranked but risky NFT project is \textsc{Cybertopia} \cite{cybertopia}. We obtain 33,413 (13.3\%) metadata from a total of 251,001 NFTs. The collected metadata contain no valid information that can represent the identity of NFT, but only duplicate data. The collectible \textsc{Sorare} \cite{sorare}, \textsc{Codex Record} \cite{codex}, and \textsc{Crypto Stamp}'s \cite{Cryptostamp} metadata are fully retrieved, where the assets are saved in \textsc{Sorare}'s server, \textsc{AWS}, and \textsc{Crypto Stamp}'s servers respectively.

        \begin{table*}[htbp]
        \resizebox{\textwidth}{!}{
        \begin{tabular}{@{}lrrrrlrrrr@{}}
        \toprule[1.5pt]
        \multicolumn{5}{l}{\textbf{Contract Aggregation}}  & \multicolumn{5}{l}{\textbf{NFTs Aggregation}}    \\ \cmidrule(r){1-5} \cmidrule(l){6-10}
        \textbf{Location} & \multicolumn{2}{r}{\textbf{\# of Contract (\%)}} & \multicolumn{2}{r}{\textbf{\# of NFTs (\%)}} & \textbf{Location} & \multicolumn{2}{r}{\textbf{\# of Contract (\%)}} & \multicolumn{2}{r}{\textbf{\# of NFTs (\%)}} \\ \cmidrule(r){1-5} \cmidrule(l){6-10}
        api.opensea.io                & 1,051            & 17.94            & 20,115          & 0.42         & api.godsunchained.com & 5               & 0.09             & 613,328         & 12.77        \\
        api.niftygateway.com          & 960             & 16.38            & 160,800         & 3.34         & amazonaws.com         & 61              & 1.04             & 496,512         & 10.34        \\
        metadata.mintable.app         & 906             & 15.47            & 25,310          & 0.53         & api.codexprotocol.com & 1               & 0.01             & 492,904         & 10.27        \\
        coludfunctions.net            & 529             & 9.03             & 41,142          & 0.86         & api.sorare.com        & 2               & 0.03             & 323,874         & 6.74         \\
        locksmith.unlock-protocol.com & 143             & 2.44             & 2,740           & 0.05         & crypto.post.at        & 8               & 0.14             & 279,047         & 5.81         \\
        herokuapp.com                 & 105             & 1.79             & 72,336          & 1.51         & cyberxtopia.com       & 1               & 0.01             & 251,001         & 5.23         \\
        api2.cargo.build              & 99              & 1.69             & 110,947         & 2.31         & api.niftygateway.com  & 960             & 15.47            & 160,800         & 3.35         \\
        azurewebsites.net             & 63              & 1.08             & 48,279          & 1.01         & blockchaincuties.com  & 1               & 0.01             & 125,173         & 2.61         \\
        amazonaws.com                 & 61              & 1.04             & 496,512         & 10.34        & mcp.town              & 9               & 0.15             & 123,072         & 2.56         \\
        factory.chocomint.app         & 60              & 1.02             & 747            & 0.01         & api2.cargo.build      & 99              & 1.69             & 110,947         & 2.31         \\ \cmidrule(r){1-5} \cmidrule(l){6-10}
        \textbf{Total \# of Contract} &&&& 5,858   &\textbf{Total \# of NFT} &&& \multicolumn{2}{r}{4,801,699}\\ \bottomrule[1.5pt]
        \end{tabular}}
        \caption{Cluster of contracts and NFTs on the off-chain servers.}
        \label{netloc}
        \end{table*}


\section{The Accessibility} 
\label{sec:accessibility}
    Given that the NFT-to-Assets path may include several layers of connectivity and storage platforms (as seen in Figure~\ref{nft2asset}), the failure of the NFT value will be caused directly by the inaccessibility of data at any point. In this section, we evaluate the availability of NFT assets by analyzing the completeness of data collection.

    \subsection{Accessibility at each stages}
    \label{sec:Accessibilityateachstages}
        The NFT-to-Assets connection is made up of four interlocking parts: \textit{TokenURI}, \textit{Metadata}, \textit{AssetURI}, and \textit{Asset}. The NFT data lost of the above four parts in the NFT contract is thereby represented by the validity rate of three transition stages, which is denoted by: \ding{182} \textit{TokenURI} $\to$ \textit{Metadata}; \ding{183} \textit{Metadata} $\to$ \textit{AssetURI}; \ding{184} \textit{AssetURI} $\to$ \textit{Asset}. The access rate is calculated by the proportion of data that is successfully retrieved from its predecessor. For example, the NFT access rate of stage \ding{182} refers to the number of collected metadata divided by the number of \textit{TokenURI}s.
        
\begin{figure}[htbp]
            \centering
            \includegraphics[width=8.5cm]{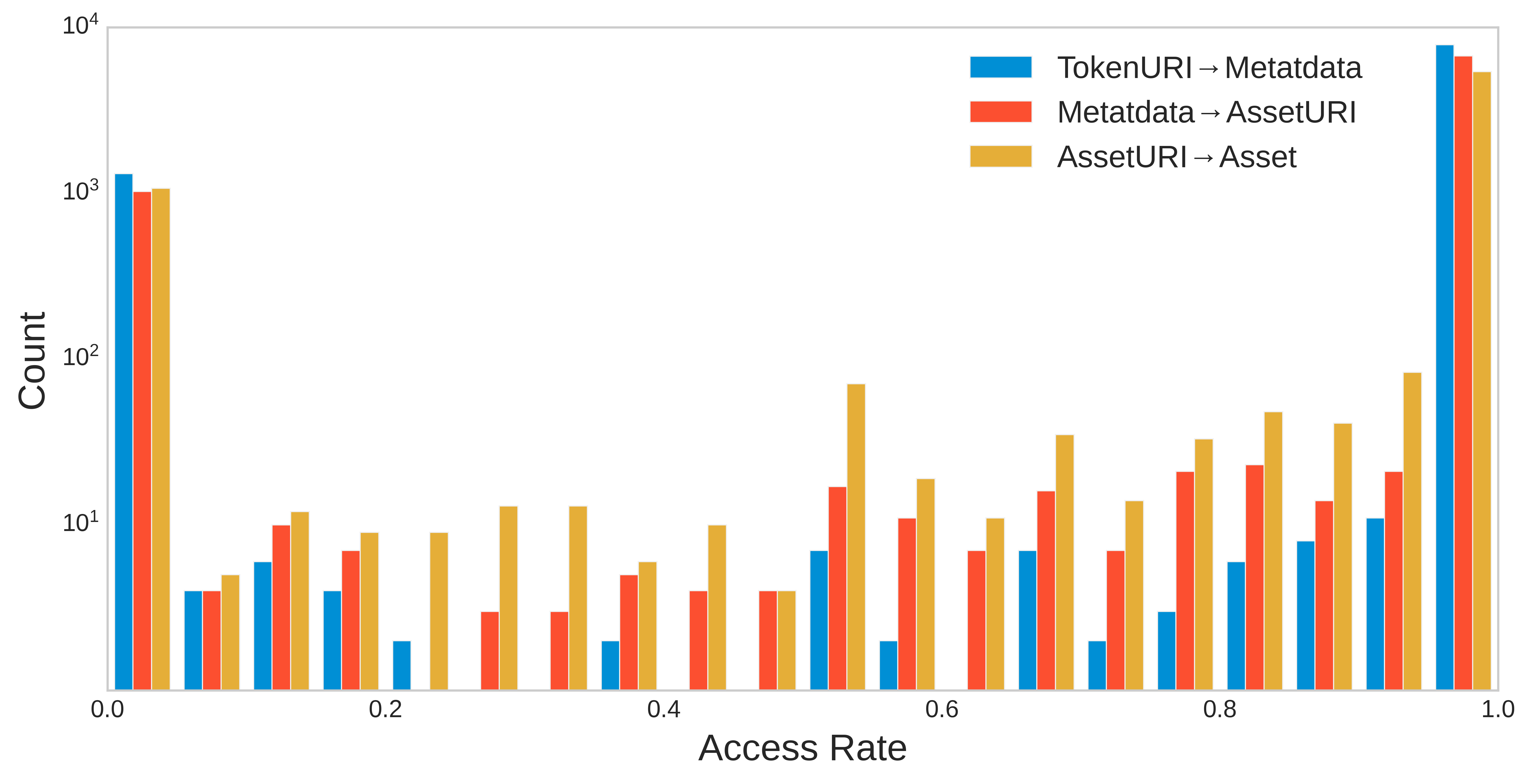}
            \caption{The distribution of access rate. \textit{TokenURI} $\to$ \textit{Metadata}, \textit{Metadata} $\to$ \textit{AssetURI}, \textit{AssetURI} $\to$ \textit{Asset} refer to \ding{182}-\ding{184} in \cref{sec:Accessibilityateachstages}.}
            \label{accessrate}
        \end{figure}

        In general, of the 9,235 NFT collections with \textit{TokenURI} retrieved from Ethereum, the number of NFT collections which is not fully retrieved is almost doubling between stages, which is 1,421 (15.39\%) of \ding{182}, 2,594 (28.09\%) of \ding{183}, and 4,004 (43.36\%) of \ding{184} respectively. As presented in Figure~\ref{accessrate}, the distribution of access rate in each stage exhibits different shapes. First, the access rate of each stage is mainly concentrated near $1.0$, indicating that \emph{most NFTs remain valid}. The number of fully valid NFTs is 7,814 (84.61\%)  of \ding{182}, 6,641 (71.91\%) of \ding{183}, and 5,231 (56.64\%) of \ding{184} respectively, while it shows \emph{an accelerated decline trend towards the end of the connection}. Secondly, per Figure~\ref{accessrate} the distribution of \ding{182} shows a tendency of more at both ends and less in the middle, which indicates that the access rate of NFTs is concentrated around $1.0$ and $0.0$, \textit{i.e.}, \emph{most NFTs are either completely available or completely unavailable}. The unavailability of \textit{TokenURI} may be due to the fact that the data obtained from \textit{TokenURI} is not itself in the form of a URL (see \cref{sec:DoseNFTpreferdecentralizedstorage}). This may also indicate that the publisher has shut down the service, since metadata is typically held by the publisher for NFTs stored on the Server. The access rate from \textit{AssetURI} to \textit{Asset} is more diffuse, with a higher percentage of NFTs in the middle interval of the figure. Such changes indicate that asset storage is much more unstable. This instability is not due to the human factor such as a service close, but rather because a large portion of assets are hosted on fragile storage platforms with poor data persistence, which is of vital importance for NFT safety. The next section expands further on the causes of data storage and retrieval failures.
        
    \subsection{How is the lost data?}
        Two perspectives are used to analyze the NFT lost data in further detail: the causes of the NFT loss and the features of these missing data. Intuitively, the NFT data lost may be potentially pertinent to the stability of the storage platform, the asset file size, the URL format, etc. To verify the factor that leads to the NFT loss, we divide the interval of NFTs' access rate into three intervals and compared the influence of the above three factors on the NFT data loss. The results are shown in Table~\ref{nftlostfactor}.
        
        \begin{enumerate}
            \item \textbf{Storage Platform.} Per Table~\ref{nftlostfactor}, the percentage of completely lost NFTs, \textit{i.e.}, $\alpha$=0, among centralized storage and decentralized storage is 53.16\% \textit{vs.} 46.84\%, which shows no significant difference between storage types. For NFTs with incomplete requests and full acquisition, the ratio of centralized storage to decentralized storage increases strikingly, about three times when 0<$\alpha$<1, and twice when $\alpha$=1. Meanwhile, as described in \cref{sec:DoseNFTpreferdecentralizedstorage}, considering that the NFTs' ratio of choosing centralized storage (5,862, 71.08\%) to decentralized storage (2,385, 28.92\%) is about 2.45, this is roughly consistent with the ratio of 0<$\alpha$<1 by twice and three times. Therefore, the proportion of the centralized NFTs with $\alpha$=0 is actually below this ratio, which indicates that \emph{the assets of the centralized storage currently have superior persistence and accessibility to some extent}.
            
            \item \textbf{Asset Size.} There is no asset size for NFTs with a 0 access rate. NFTs with 0<$\alpha$<1 are characterized by a large number of NFTs within contracts, nearly 1.77 M NFTs in 843 contracts, which poses challenges to maintaining asset integrity. However, the average size of these assets is only 1.06 MB, which is slightly smaller than the NFT assets with $\alpha$=1. We thereby conduct a two-sample $t$-test on the average asset size of the set of 0<$\alpha$<1 and $\alpha$=1 to test whether the means of the two independent samples are significantly different. The test is two-sided and the null hypothesis $H_0$ assumes that the means of the two samples are the same. Results show that the statistic $t$=-2.846 and the \textit{p-value}=0.004. As the \textit{p-value} is smaller than 0.025, the null hypothesis can be rejected with 95\% confidence, and there exist significant differences between the average asset size. Therefore, \emph{the average file size of NFTs with $\alpha$=1 is indeed larger than that of 0<$\alpha$<1}.
            
            \item \textbf{URL Format.} The number of non-retrieval \textit{AssetURI} formats are presented. For NFTs with $\alpha$=0, 2,653 (78.44\%) of asset access failures are caused by illegal URLs, and the primary cause for this failure is that the return value of these URLs is null. The influence of invalid URLs is tiny for NFTs of $\alpha\neq$1, and only a few of these NFTs are unreachable due to empty URLs.
        \end{enumerate}
        
        \begin{table*}[htbp]
    \resizebox{\textwidth}{!}{  
        \begin{tabular}{@{}crrrrrrrrr@{}}
        \toprule[1.5pt]
        \multirow{2}{*}{\textbf{\makecell{Interval}}} & \multicolumn{1}{c}{\multirow{2}{*}{\textbf{\makecell{\# of \\Contract}}}} & \multicolumn{1}{c}{\multirow{2}{*}{\textbf{\makecell{\# of \\NFT}}}} & \multicolumn{2}{c}{\textbf{Storage}}   & \multicolumn{2}{c}{\textbf{File Size}}   & \multicolumn{3}{c}{\textbf{URL Format}}   \\ \cmidrule(r){4-5} \cmidrule(r){6-7} \cmidrule(l){8-10}
        & \multicolumn{1}{c}{}    & \multicolumn{1}{c}{}   & \multicolumn{1}{c}{\textbf{IPFS + AR}} & \multicolumn{1}{c}{\textbf{Server}} & \multicolumn{1}{c}{\textbf{Overall}} & \multicolumn{1}{c}{\textbf{Average}} & \multicolumn{1}{c}{\textbf{Empty}} & \multicolumn{1}{c}{\textbf{Non-URL}} & \multicolumn{1}{c}{\textbf{Asset}} \\ \cmidrule(r){1-3} \cmidrule(r){4-5} \cmidrule(r){6-7} \cmidrule(l){8-10}
         $\alpha$ = 0  & 3,382   &   $\backslash$   & 226 + 18 (53.16\%)   & 215 (46.84\%)  & $\backslash$    & $\backslash$  & 2,746   & 28   & 149    \\
         0 < $\alpha$ < 1 & 843    & 1,773,287    & 190 + 27 (25.93\%)    & 620 (74.07\%)  & 1.80 TB   & 1.06 MB  & 5  & 0    & 0   \\
         $\alpha$ = 1  & 5,010     & 1,130,648    & 1,360 + 331 (33.77\%)   & 3,317 (66.23\%)  & 1.50 TB  & 1.39 MB   & 2  & 0   & 0   \\ \bottomrule[1.5pt]
        \end{tabular}
        }
        \caption{The influence of storage platform, asset volume and other factors on the NFT preservation. The arrival rate of \textit{TokenURI} towards \textit{Assets} is represented by $\alpha$.}
        \label{nftlostfactor}
        \end{table*}

\section{The Assets}
\label{sec:assets}
    In this section, we take a first look at the landscape of the content formats and compare their preference across various storage platforms. we study the NFT plagiarism from the perspective of NFT duplication, where the assessment target composes of the repeated URLs and the repeated underlying asset, aiming to locate the exact phase where duplication massively occurs.
    
    \subsection{Content Format}
        The essence of NFTs is a reflection of diverse real-world data onto the blockchain (\textit{e.g.}, images, videos, texts, attribute table). In the absence of relevant data, NFTs are merely a meaningless identifier in the code of the smart contract. The primary status of various asset formats amongst NFTs is illustrated in Table~\ref{typeandsize}. The assets' file types are recognized via python \textit{Magic} modules \cite{magic}.
         
        According to the result, there are 24 asset types, and the aggregated volume of 2,851,894 NFT assets is approximately 3.29TB, of which \textit{Image}s account for about 84.61\% and \textit{Media} for 15.30\%. Considering that the hit ratio of the collected assets from a total number of NFTs is accordingly 45.75\% (Table~\ref{NFTdata}), we estimate that the entire NFT ecosystem needed additional storage space of 7.19 TB to fully save the NFT underlying assets at present. We find that NFTs in image format occupy about 84.61\% of the total storage and specifically the \textit{.png} image takes up 58.48\% of NFTs.
        
        We make a further assessment to identify whether the location of storage (\textit{i.e.}, decentralized or centralized) affects the format or size of the NFT asset. To showcase, the primary types are selected and are presented in Table~\ref{typeandsize_diff}. It gives that the centralized and decentralized hosted assets are respectively 1.18TB (35.87\%) and 2.11TB (64.13\%) in size, as well as 325,395 (11.41\%) and 2,526,499 (88.59\%) in row count. The average asset size stored in decentralized platforms is 4.34 times that of centralized ones, indicating that \emph{larger data inclines to store on decentralized platforms}. 

        \begin{table*}[htbp]
        \resizebox{\textwidth}{!}{
        \begin{tabular}{@{}lrrlrrlrrlrrlrr@{}}
        \toprule[1.5pt]
        \multicolumn{6}{c}{\textbf{Image}} & \multicolumn{3}{c}{\textbf{Media}} & \multicolumn{3}{c}{\textbf{Webpage}}  & \multicolumn{3}{c}{\textbf{Others}} \\ \cmidrule(r){1-6} \cmidrule(rl){7-9} \cmidrule(rl){10-12} \cmidrule(l){13-15}
        \textbf{Ext} & \textbf{Count} & \textbf{Size} & \textbf{Ext} & \textbf{Count} & \textbf{Size} & \textbf{Ext} & \textbf{Count} & \textbf{Size} & \textbf{Ext} & \textbf{Count} & \textbf{Size} & \textbf{Ext} & \textbf{Count} & \textbf{Size} \\ \cmidrule(r){1-6} \cmidrule(rl){7-9} \cmidrule(rl){10-12} \cmidrule(l){13-15}
        .png  & 1,667,844 & 1.09TB  & .webp & 2,272                & 2.32GB             & .mp4 & 71,637               & 513.28GB                     & .xpdl                & 12,488               & 2.95MB               & .wasm  & 4,756  & 2.80GB   \\
        .jfif & 547,157   & 342.06GB & .tif  & 26                   & 4.65GB               & .mov & 421                  & 2.30GB                       & .htm                 & 2,491                & 2.01MB               & .pdf   & 3      & 59.42MB  \\
        .jpeg & 207,532   & 425.12GB & .tiff & 5                    & 79.26MB              & .wav & 1                    & 23.33MB                      & .xhtml               & 83                   & 2.67MB             & .pl    & 2      & 51.00B   \\
        .svg  & 183,947   & 10.73GB  & .glb  & 3                    & 15.41MB              & .mp2 & 7                    & 2.80MB                       & .xml                 & 3                    & 438.00B              & .so    & 2      & 78.06KB  \\
        .gif  & 121,037   & 944.37GB & .webm & 1                    & 5.78MB               &      & \multicolumn{1}{l}{} &                              & \multicolumn{1}{r}{} & \multicolumn{1}{l}{} & \multicolumn{1}{l}{} & .bin   & 1      & 11.22MB  \\
        .jpg  & 25,533    & 8.15GB   &       & \multicolumn{1}{l}{} & \multicolumn{1}{l}{} &      & \multicolumn{1}{l}{} &                              &                      &                      &                      &        &        &          \\ \cmidrule(r){1-6} \cmidrule(rl){7-9} \cmidrule(rl){10-12} \cmidrule(l){13-15}
        \multicolumn{5}{l}{\textbf{Sum of Category}} & 2.78TB &  &  & 515.61GB &  &  & 7.62MB  &  &  & 2.87GB   \\ \bottomrule[1.5pt]
        \end{tabular}}
        \caption{File type and file size of NFT corresponding assets.}
        \label{typeandsize}
        \end{table*}
        
    
        \begin{table*}[htbp]
        \resizebox{\textwidth}{!}{
        \begin{tabular}{@{}lrrrrrrrrrrrr@{}}
        \toprule[1.5pt]
              & \multicolumn{12}{c}{\textbf{Centralized Storage: Server}}   \\ \midrule
              & \textbf{.png}      & \textbf{.jfif}    & \textbf{.jpeg}    & \textbf{.svg}     & \textbf{.gif}     & \textbf{.jpg}     & \textbf{.webp}    & \textbf{.tif}   & \textbf{.tiff}    & \textbf{.mp4}     & \textbf{.mov}     & \textbf{.wasm}  \\ \cmidrule(l){2-13}
        \textbf{Size}  & 794.46GB  & 247.21GB & 294.23GB & 10.87GB  & 570.62GB & 7.30GB   & 549.93MB & 4.65GB & 44.92MB & 226.87GB & 1.48GB   & 2.80GB \\
        \textbf{Count} & 1,525,582 & 509,221  & 153,669  & 182,768  & 51,241   & 25,296   & 2,051    & 26     & 3       & 56,400   & 390      & 4,756  \\ \midrule
              & \multicolumn{12}{c}{\textbf{Decentralized Storage: IPFS + Arweave}}  \\ \midrule
              & \textbf{.png}      & \textbf{.jfif}    & \textbf{.jpeg}    & \textbf{.svg}     & \textbf{.gif}     & \textbf{.jpg}     & \textbf{.webp}    & \textbf{.tif}   & \textbf{.tiff}    & \textbf{.mp4}     & \textbf{.mov}     & \textbf{.wasm}  \\ \cmidrule(l){2-13}
        \textbf{Size}  & 318.89GB  & 94.86GB  & 130.89GB & 161.14MB & 373.75GB & 873.01MB & 1.79GB   & 0.00B  & 34.34MB  & 286.41GB & 835.31MB & 0.00B  \\
        \textbf{Count} & 142,262   & 37,936   & 53,863   & 1,179    & 69,796   & 237      & 221      & 0      & 2        & 15,237   & 31       & 0      \\ \bottomrule[1.5pt]
        \end{tabular}}
        \caption{Storage preferences of different file types.}
        \label{typeandsize_diff}
        \end{table*}
        
    
    \subsection{Duplicated URL}
    \label{sec:duplicatedurl}
\begin{figure*}[htbp]
            \centering
            \includegraphics[width=\textwidth]{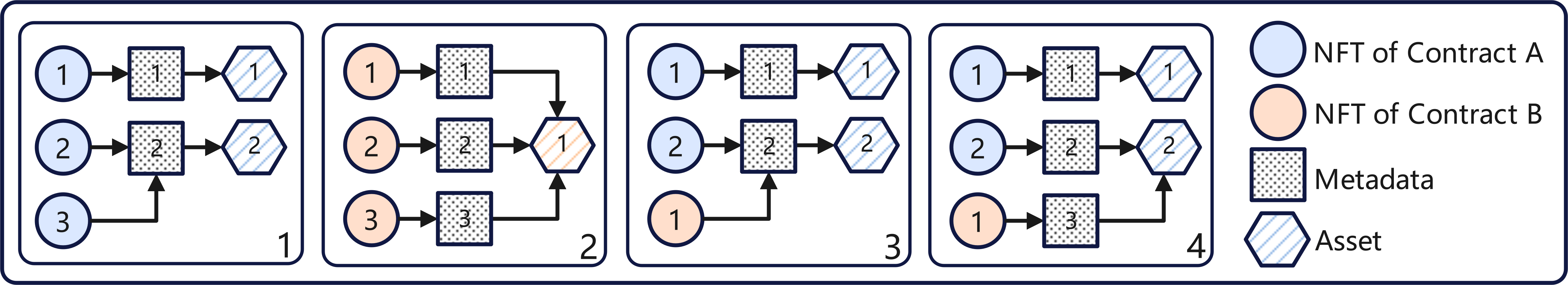}
            \caption{Types of URL confusion.}
            \label{pointer}
        \end{figure*}

        We observe that a large number of NFT collections have encountered the problem of high duplication of corresponding URLs. NFTs may share the same metadata URL, and consequently have identical metadata and asset data, which compromises the uniqueness property of NFT. Considering that a large proportion of NFTs adopt two-stage storage, such repetitiveness can thus be generalized into 4 fundamental situations, which is presented in Figure~\ref{pointer}: \ding{182} NFTs within a same contract have identical \textit{TokenURI}; \ding{183} NFTs within a contract have identical \textit{AssetURI}; \ding{184} NFTs of different contracts have identical \textit{TokenURI}; \ding{185} NFTs of different contracts have identical \textit{AssetURI}. To understand the extent to which NFTs suffer from representational non-uniqueness, we make an assessment of the above situations respectively and the complex effect caused by them. We study the duplicated URL in terms of the above fundamental types.
        
        \subsubsection{Type \textbf{(1)}} \label{sec:type1} A straightforward approach to measure the duplicated rate of the NFT corresponding URLs is to identify the percentage of the URLs that are redundant. For instance, suppose that a contract has 10 NFTs, and the NFTs' \textit{TokenURI} series is denoted by \{ \textit{A,A,A,A,B,B,B,C,C,D} \}. The 10 URLs can be then deduplicated to 4 URLs, which is A, B, C, and D, while the rest 6 URLs are redundant. Then the duplicated rate is calculated as the number of redundant URLs divided by the total number of URLs, which is 60\%. In this case, we observe that 1,756 (23.50\%) NFT contracts share \textit{TokenURI} to varying degrees. \emph{The repetition rate of NFTs with repeated URLs is mainly concentrated toward 100\%}. As can be seen from Figure~\ref{uriduplication}, the number of NFTs is highly concentrated when the repetition rate is higher than 50\%. Of these, \emph{527 (30.01\%) contracts have a repetition rate higher than 95\%}. Part of the problem with these ultra-high repetition rate NFTs is that the \textit{TokenURI} of the NFT returns a value that is either \textit{Null} or \textit{None}.
        
        \subsubsection{Type \textbf{(2)}} \label{sec:type2} The calculation of duplicated \textit{AssetURI} may be inaccurately estimated for the following reasons: \ding{182} The \textit{AssetURI} usually includes multiple field for recording URLs, rendering different duplicated rates of each field of URLs, \textit{i.e}., \textit{image}, \textit{external\_url} (refer to \cref{sec:assetcollection}).; \ding{183} String of metadata URL field is the URL base shared between NFTs within the NFT contract, while the \textit{AssetURI} is already included in other fields, resulting in an overestimation of the duplicated rate. To address the aforementioned problems, the duplicated rates of each field of URLs are calculated separately, and the aggregate duplicated rate is the minimum value of the duplicated rates in fields. Therefore, the duplicated rate is determined by the lowest repetition rate of URLs in different fields.
        
\begin{figure}[htbp]
            \centering
            \includegraphics[width=8.5cm]{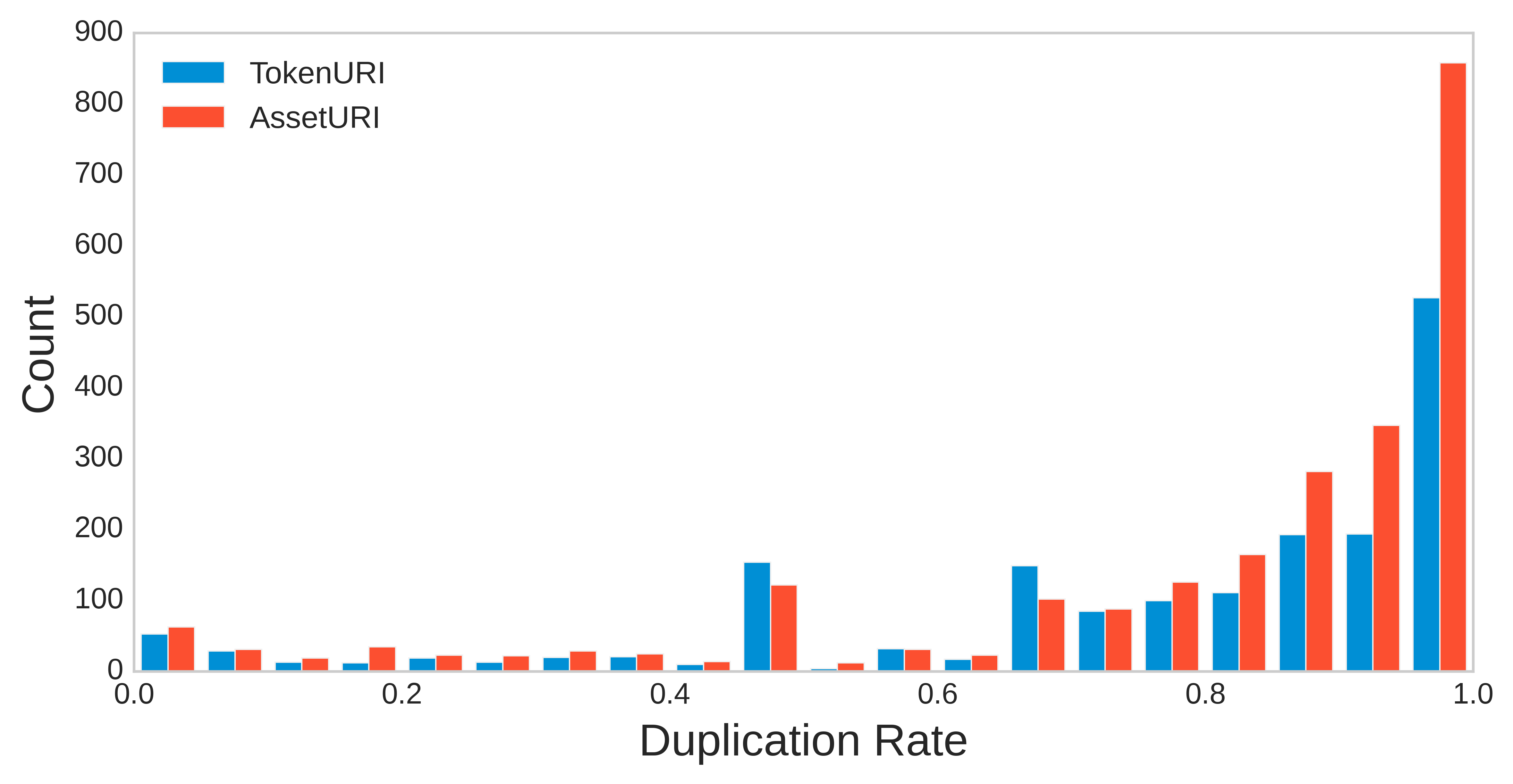}
            \caption{The distribution of duplicated rate of NFTs' \textit{TokenURI} (\cref{sec:type1}) and \textit{AssetURI}(\cref{sec:type2}), excluding cases where URIs are not duplicated.}
            \label{uriduplication}
        \end{figure}

        According to the results, the \textit{AssetURI}s are \emph{severely} duplicated. \emph{There exists 2,417 (37.95\%) NFT collections that have duplicated \textit{AssetURI} NFTs, and 890 (36.82\%) redundant NFT collections have duplicated rate larger than 95\%}. Per Figure~\ref{uriduplication} the number of NFT collections with repeated \textit{AssetURI} increased significantly compared to \textit{TokenURI}. Furthermore, the 661 newly included duplicated NFT collections largely piled up in the range where the duplicated rate is larger than 75\%. \emph{This phenomenon indicates a decoupling between NFT's on-chain metadata and off-chain assets, as a large number of different \textit{TokenURI}s refer to identical \textit{AssetURI}}.

        \subsubsection{Type \textbf{(3)}} \label{sec:type3} Of this type, we detect 26,983 \textit{TokenURI}s which are shared among 528 contracts, and the most widely shared URL is used by 15 contracts. We then count the addresses that occurred most frequently in all shared contract groups, which is the number of other contracts that has \textit{TokenURI} overlapped with it, aiming to observe the association of shared addresses between contracts. We find that the the most involved NFT has \textit{TokenURI} overlapped with 17 other NFTs. However, the \textit{TokenURI} of NFTs with large amount of shared contracts (the top-10 Contracts) are unusable information (numbers, invalid URLs, redirected URLs). And common characteristics of the shared-URL contracts is that most of them have the same contract name, and it probably includes counterfeit contracts, or the contracts are repeatedly deployed for project testing or release updates. It suggests that \emph{the \textit{TokenURI} duplication is more of a duplication of invalid information and multiple contracts belonging to a same project.}
        
        \subsubsection{Type \textbf{(4)}} \label{sec:type4} We observe 23,340 (0.64\%) URLs shared by NFTs of different contracts out of 3,652,097 \textit{AssetURI}s, for a total of 491 such contracts. Meanwhile, there are 788 combinations of NFT contracts that share the same URL, and the number of shared URLs of each contract combination ranges from 1 to 10,000, but 98.10\% of contract groups have less than 100 shared links. As with Type \textbf{(3)}, the addresses with the highest number of shared URLs tend to have similar token names. For example, the contract group with 10,000 shared URLs is made up of 4 addresses named \textsc{Su Squares} (\texttt{SU}) \cite{SuSqares}, and the contract group that shares 4,606 URLs is made up of 3 addresses named \textsc{89 seconds Atomized} (\texttt{SNP001}) \cite{secondsAtomized}. The relevance among contracts is not heavily associated. The most severe situation is that one contract shared the URLs with 15 other contracts, while 89.21\% of the contracts are shared with only one other contract. That is, \emph{the vast majority of \textit{AssetURI} cross-contract sharing is taking place on a tiny scale}.

    \subsection{Identical Underlying Content}
    \label{sec:identicalcontent}    
        Assets obtained from different URLs may still probably identical. Analysis of duplicate URLs alone is insufficient to illustrate the degree of the duplication of assets. The identical assets are detected using hash function at first (\textit{e.g.}, \texttt{MD5}, \texttt{SHA256}). Whereas the asset may be modified and used for minting new NFT, the numerical different content may probably be visually similar with the aforementioned hash function. These hash functions yields different results against such minor change, which is unable to detect the nearly identical content. Therefore, in the second step, a similarity hash algorithm is applied to measure the extent to which assets are duplicated (\textit{e.g.}, \texttt{dHash}, \texttt{aHash}, \texttt{pHash}), which provides a fault tolerant measurement on the asset similarity.

        \subsubsection{Asset Hash}
        \label{sec:assethash}
\begin{figure}[htbp]
            \centering
            \includegraphics[width=7cm]{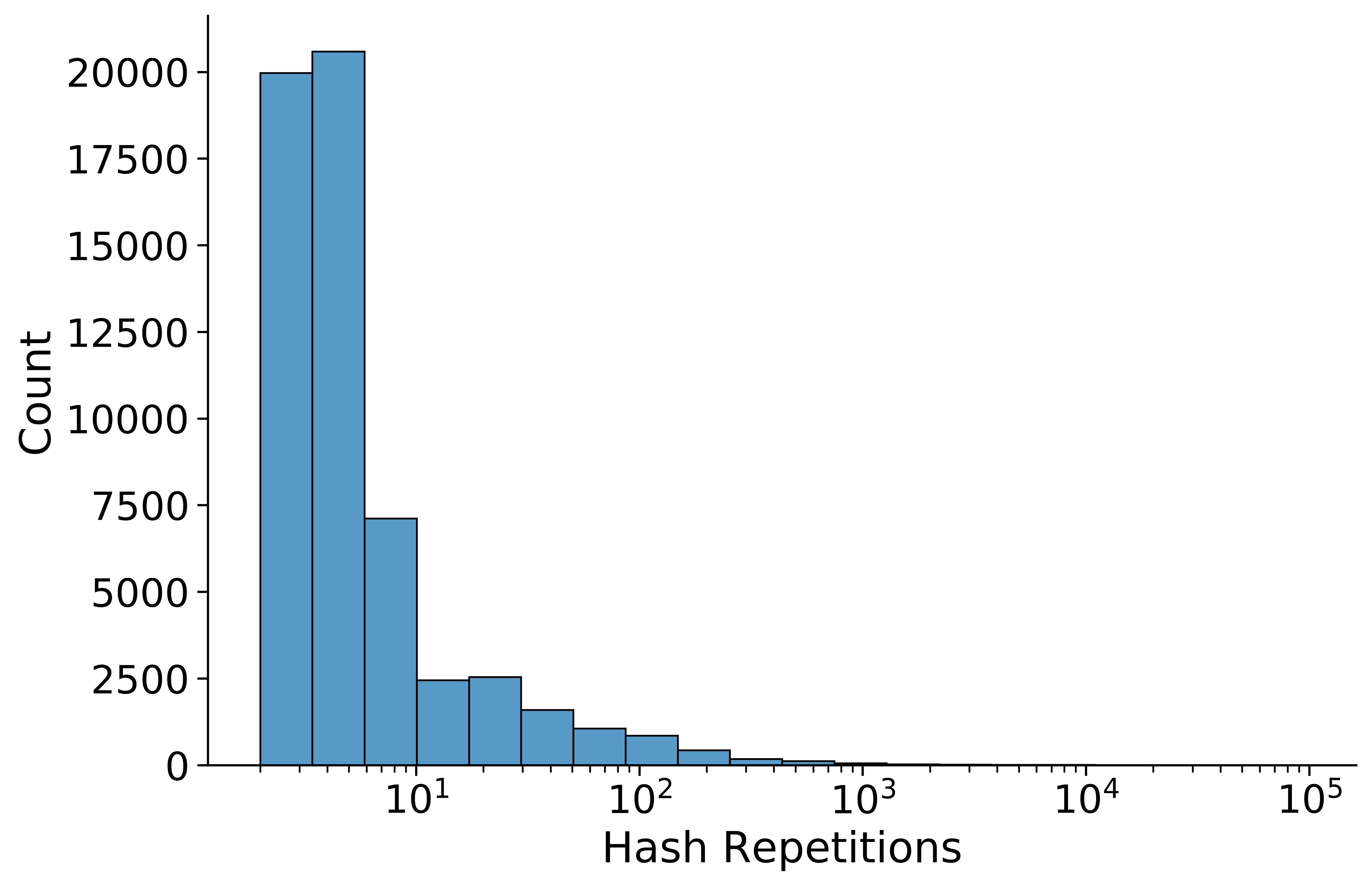}
            \caption{The distribution of repetition of duplicated hash in log scale.}
            \label{assethash_repetition}
        \end{figure}

        The quantitative results are presented in Table~\ref{hashresult}. The duplication rate under the measure of asset hash is consistent with the method stated in \cref{sec:type1}. The number of duplicated NFT collections increases from 2,417 (at \textit{AssetURI} level in \cref{sec:type2}) to 2,653. In other words, \emph{the NFTs of 236 collections exist duplicated assets when there is no duplication of \textit{AssetURI} (i.e., identical content from different AssetURIs)}. The average duplication rate within these NFT collections is 34.68\%. As a whole, there are a total number of 1,412,662 unique hashes of the assets, of which 95.96\% has a frequency equal 1 that these NFT assets have no replica, indicating that \emph{the majority of the NFT contents are still unique}. The distribution of repetition of the rest of the 57,079 replicated content hashes are presented in Figure~\ref{assethash_repetition}. Per this plot the bars almost vanish above $10^{3}$ and 109 hashes that are repeated more than $10^{3}$. The most repeated hash is the image sign of ``\textit{MOVED TO MATIC}'', which comes from the blockchain game \textsc{MegaCryptoPolis} \cite{mcp3d} and is shared by 94,749 NFTs of 6 contracts. We observe that \textsc{MegaCryptoPolis} has deployed these 6 NFT contracts to Ethereum and there are 100,814 NFTs in sum. The assets of \textsc{MegaCryptoPolis}'s NFTs used to be hosted on application providers' servers in image format, but are now migrated to \textsc{Matic Network} \cite{matic}, and hence the original NFTs on Ethereum lost their value. This also shows that the owner of the NFTs is not necessarily its custodian.
        
        Furthermore, we find that 12,688 duplicated assets are identically owed by multiple contracts and 44,391 of that are duplicated within a single contract. It suggests that \emph{77.77\% of the duplication occurs within an NFT collection}. The average repetitions of duplicated assets among multiple contracts and within a contract are 28.24 and 25.63 respectively, which shows no significant difference. We then verify the most associated contracts empirically, and we observe that a large proportion of the contract groups that share identical content are also correlated in the contract names and symbols. They may contain counterfeit contracts, or they may be different contracts published by application providers, which is in line with the results of duplicated URLs (refer to \cref{sec:type3}, \cref{sec:type4}). Table~\ref{caseofsharecontent} shows a case of a group of \texttt{PGFK} contracts \cite{pgfk} which shares identical image. The sharing of this image occurs both within individual contracts and between similar contracts. 
        
        \begin{table}[ht]
        \resizebox{8.5cm}{!}{
        \begin{tabular}{@{}llrr@{}}
        \toprule[1.5pt]
        \textbf{Contract Name}                               & \textbf{Symbol}   & \multicolumn{1}{l}{\textbf{\makecell{Address of \\1st 10-digits}}} & \multicolumn{1}{l}{\textbf{\makecell{No. of \\ Repetitions}}} \\ \midrule
        PGFKs 26                           & PGFK 26  & 0xae0ae230                                & 26                                 \\
        PGFKs 181                          & PGFK 181 & 0x46baaa89                                & 181                                \\
        PoIyient Games Founders Keys 41    & PGFK 41  & 0xb56fe48a                                & 41                                 \\
        Polyient Game Founders Keys 25     & PGFK 25  & 0xa8a0211b                                & 25                                 \\
        Polyient Game Founders Keys 25     & PGFK 25  & 0x51479ee7                                & 25                                 \\
        Polyient Games Founders Keys ?? 10 & PGFK 10  & 0x773da030                                & 10                                 \\ \midrule
        Shared Image Hash (\texttt{MD5})          & \multicolumn{3}{r}{b05f8083ca4d68f423fd61515cd9a803}                                      \\ \bottomrule[1.5pt]
        \end{tabular}}
        \caption{A case of specific content that is shared among 6 NFT contracts.}
        \label{caseofsharecontent}
        \end{table}


        \subsubsection{Image Fuzzy Hash} 
            The widely-used hash algorithm \textit{Perceptual Hash} (\texttt{pHash}) and \textit{Block Mean Value based Hash} (\texttt{BlockHash}) are selected to measure the situation of NFT image similarity \cite{hao2021s}. For \texttt{pHash}, we use an open-source library \texttt{ImageHash} for the experiment \cite{ImageHash}; For \texttt{BlockHash}, we use another open-source implementation \texttt{blockhash} for the experiment \cite{BlockHash}. The images are resized to $256*256$ at first and are then applied to the chosen hash algorithm. The precision of fuzzy hash is controlled by the hash size range in [$8, 16, 32, 64, 128, 256$] bits, as it determines the granularity of image segmentation. In general, the finer the granularity of the image segmentation, the more distinct  the images will be, resulting in lower repeatability.
            
\begin{figure}[htbp]
                \centering
                \includegraphics[width=11cm]{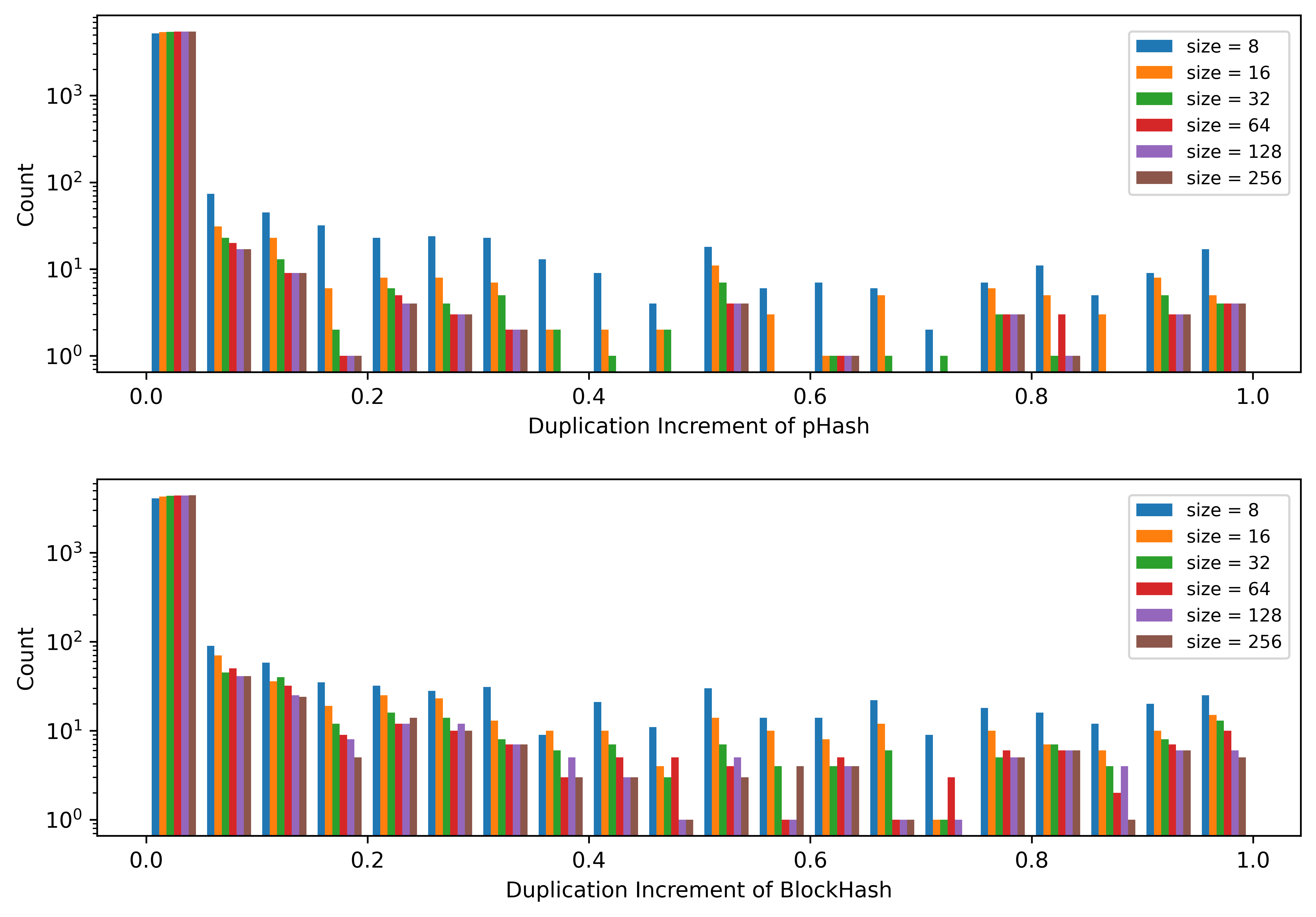}
                \caption{The duplication increment with perceptual hash on the base of conventional Hash, where \textit{size} refers to the length of hash string.}
                \label{hash_incre}
            \end{figure}

            We measure the degree of NFT images similarity with the tolerance of small perturbations by comparing the result of the perceptual hash to the conventional hash (the \texttt{MD5} results used in \cref{sec:assethash}). The result is explained as the increase in duplication, which is the duplication under perceptual hash minus the duplication under ordinary hash, where the calculation of duplication is again consistent with the method stated in \cref{sec:type1}. Notice that the measurement scope is narrow to the pure image NFT collection as the perceptual hash is limited to image content, as the perceptual hash can only be applied to image, which will cause a negative increment to the mixed-type NFTs. The distribution of duplication increment of NFT collections is presented in Figure~\ref{hash_incre}. The increase of the vast majority of NFTs is still near 0 and the bars sharply decrease to the right, indicating that \emph{the situation of similar content is not severe yet}. Another observation is that the NFTs clustering in the middle ($0.4 \sim 0.8$) are sparser than at the right end, which suggests that the nearly identical NFTs are not uniformly distributed, but is more likely due to the original setting of the NFT collection (\textit{e.g.}, monochrome background, same shape in different colors).
            
            The numerical results of the perceptual hash are also aggregated in Table~\ref{hashresult}. As the size of the perceived hash increases, the number of duplicate assets tends to the conventional hash algorithm. When the hash size is 8, the number of duplicate assets of \texttt{pHash} and \texttt{BlockHash} is 3.37\% and 6.55\% higher than that of the original hash algorithm, respectively. However, when the hash size is greater than 32, the difference in similarity tends to be negligible. \texttt{pHash} in general is more sensitive to asset similarity, and the number of duplicate assets increases almost linearly, whereas \texttt{BlockHash} increments with small hash sizes are much larger than its increments with large hash sizes. \textit{The proportion of the duplicate assets within a single contract versus across multiple contracts are all around 99\% and 1\% with variance no greater than 0.9\%}. So in this sense, neither \texttt{pHash} nor \texttt{BlockHash} shows a significant difference from \texttt{MD5}. Nevertheless, the average frequency of the multi-contract hash of \texttt{BlockHash} is considerably larger than that of \texttt{MD5} and \texttt{pHash} around 5 to 7 times, indicating that \emph{the images shared by multiple contracts are largely identical under the metric of \texttt{BlockHash}}.

            \begin{table}[htbp]
            \centering
            \resizebox{13cm}{!}{
            \begin{tabular}{@{}lrrrrrrr@{}}
            \toprule[1.5pt]
            
            \multirow{2}{*}{} & \multirow{2}{*}{\textbf{\makecell{\# of \\ Unique Hash}}} & \multirow{2}{*}{\textbf{\makecell{Prop. of \\Dup Hash}}} & \multirow{2}{*}{\textbf{\makecell{Prop. of \\Non-dup Hash}}} & \multicolumn{2}{c}{\textbf{Multi-Contract}}            & \multicolumn{2}{c}{\textbf{Single Contract}}           \\ \cmidrule(l){5-6} \cmidrule(l){7-8} 
            &     &      &     & \textbf{Proportion}  & \textbf{Frequency}  & \textbf{Proportion}  & \textbf{Frequency}   \\ \midrule
            
            \textbf{Hash} &&&&&&&\\ \midrule
            \texttt{MD5}      & 1,412,662         & 4.04\%         & 95.96\%   & 0.89\%              & 28.25         & 99.11\%             & 25.63          \\ \midrule
            \textbf{pHash} &&&&&&&\\ \midrule
            size = 8       & 1,046,644         & 7.41\%         & 92.59\%   & 0.86\%              & 41.42         & 99.14\%             & 17.93          \\
            size = 16      & 1,154,287         & 4.98\%         & 95.02\%   & 1.07\%              & 28.79         & 98.93\%             & 24.75          \\
            size = 32      & 1,188,263         & 5.40\%         & 94.60\%   & 1.03\%              & 28.62         & 98.97\%             & 21.11          \\
            size = 64      & 1,226,512         & 4.40\%         & 95.60\%   & 1.00\%              & 28.58         & 99.00\%             & 25.10          \\
            size = 128     & 1,226,742         & 4.40\%         & 95.60\%   & 1.00\%              & 28.40         & 99.00\%             & 25.15          \\
            size = 256     & 1,226,787         & 4.40\%         & 95.60\%   & 1.00\%              & 28.40         & 99.00\%             & 25.16          \\ \midrule
            \textbf{BlockHash} &&&&&&&\\ \midrule
            size = 8   & 690,782           & 10.59\%        & 89.41\%   & 0.87\%              & 77.79         & 99.13\%               & 15.95          \\
            size = 16  & 931,937           & 6.80\%         & 93.20\%   & 0.16\%              & 205.96        & 99.84\%               & 15.73          \\
            size = 32  & 998,743           & 5.19\%         & 94.81\%   & 0.16\%              & 194.54        & 99.84\%               & 17.96          \\
            size = 64  & 1,025,215         & 4.60\%         & 95.40\%   & 0.17\%              & 179.18        & 99.83\%               & 19.23          \\
            size = 128 & 1,039,439         & 4.28\%         & 95.72\%   & 0.17\%              & 170.02        & 99.83\%               & 20.09          \\
            size = 256 & 1,040,032         & 4.11\%         & 95.89\%   & 0.18\%              & 159.68        & 99.82\%               & 20.93          \\ \bottomrule[1.5pt]
            \end{tabular}}
            \caption{The similarity of NFT assets measured with conventional Hash, pHash and BlockHash. \textit{The number of Unique Hash} indicates the number of unique NFT assets. The \textit{Dup Hash}, and \textit{Non-dup Hash} refer to whether the hash corresponding asset has duplication or not. The \textit{Multi-Contract} and \textit{Single Contract} refer to whether the duplicated hash corresponding asset exists in multiple contracts or within a single contract.}
            \label{hashresult}
            \end{table}
            
            \vspace{-1mm}

\section{Related Work}
    
    
    \vspace{1mm}
    \noindent{\ding{228}} \textbf{Market Trend}. Studies on the factors that affect NFT market and how they achieve are diverse: \ding{182} \textbf{Rarity.} The rarity of NFTs is regarded as an important factor that determined their endogenous value. \citet{piyadigama2022analysis} applied a cosine similarity matrix to score on NFT traits, and the score (or scarcity) was used to evaluate the NFT value. \citet{mekacher2022rarity} analyzed whether rarity increases NFT transactions and attracts new buyers, and whether high rarity leads to higher returns. \ding{183} \textbf{Cryptocurrency.} \citet{dowling2022non} explored the relation between NFT pricing and cryptocurrency pricing, and results showed that price volatility pass-through between cryptocurrencies and NFTs is limited, but synergies do exist. \citet{ante2022non} investigated the interrelationships between NFT sales and the pricing of BTC and ETH, which suggested that cryptocurrency markets will affect the growth of the smaller NFT market, yet there is no opposite impact. \ding{184} \textbf{Social Network} \citet{casale2022impact} studied the interaction between the volume, users, and social engagement on Twitter and the transaction volume of the selected NFT. \citet{kapoor2022tweetboost} evaluated the growth of NFTs and outlines the characteristics of Twitter users who promote NFTs sales. \ding{185} \textbf{Artist.} \citet{vasan2022quantifying} studied the characteristics, mechanisms, and networks of crypto-artists to discover how to attain success as a crypto-artist. \ding{186} \textbf{Auction.} \citet{kireyev2022nft} focused on the costs of bidding to show how the marketplace affects NFT trade, which might vary among markets depending on transaction fees and the frequency of bidding bots. In addition, attempts have been made to forecast the value of NFTs using statistical techniques \cite{bao2022non}, machine learning \cite{nadini2021mapping, kapoor2022tweetboost}, and deep learning \cite{nadini2021mapping, jain2022nft, kapoor2022tweetboost}. \citet{ito2022predicting} fit NFT price into LPPL Model to predict the bubble of some specific NFT collections. Besides, \citet{ko2022economic} put forth NFT trading in portfolios and analyze the potential for higher returns in portfolio investments. \citet{casale2021networks} studied the topological of NFT networks and characterize the market influencers.

    \vspace{1mm}
    \noindent{\ding{228}} \textbf{Security and Privacy.} \citet{das2021understanding} summarized a comprehensive list of design flaws of NFT and conducted a quantitative analysis to explain the severity. \citet{von2022nft} focused on a more specific field to study the NFT wash trading transaction by detecting illicit pattern of trading graph. \citet{pungila2022new} and \citet{galis2022fast} applied a pattern matching methods to detect the plagiarism of NFT. At present, NFTs have very little anonymity and privacy, which is due to the publicity of blockchain and storage platform. The only way to protect content privacy is to encrypt the data before publishing it, as anyone can access the file via its \texttt{CID}. \citet{wang2021non} assessed the security of NFT data, including spoofing, tampering, disclosure, as well as the privacy issues of current NFT, and pointed out potential solutions. \citet{battah2020blockchain} proposed to use a group of oracle to manage the granting of multi-party authorization key of accessing encrypted data on public storage platform. \citet{songzkdet} designed \textsc{ZKDET} with zero-knowledge proof to achieve both traceability and privacy-preserving during NFT trading.

    \vspace{1mm}
    \noindent{\ding{228}} \textbf{Infrastructure.} \ding{182} \textbf{Storage.} \citet{de2021accelerating} extended and improved \textsc{BitSwap} \cite{bitswap}, the underlying protocol of \textsc{IPFS} and \textsc{Filecoin}, to achieve high efficiency of the storage network, including reducing communication overhead and increasing content discovery rate. \citet{daniel2022ipfs} gave a technical comparison of the existing P2P storage network, which includes an assortment of file systems (\textit{e.g.}, \textsc{Storj} \cite{storj}, \textsc{BitTorrent} \cite{bittorrent}, and \textsc{SAFE} \cite{safe}) that the NFT has not yet widely adopted. \ding{183} \textbf{Incentive.} \citet{chen2022absnft} proposed a safety repurchase mechanism \texttt{ABSNFT} that assures buyers' interests, where the bidding behavior in the repurchase is formulated as a game. From another perspective, \citet{hasanincorporating} proposed a rewards and punishment mechanism to regulate the NFT trading behavior. \ding{184} \textbf{NFT on \textsc{Fabric}.} \citet{bal2019nftracer} proposed \texttt{NFTracer} on \textsc{Hyperledger Fabric} to secure the verification and authentication of real-world data via the NFT technique. \citet{hong2019design} constructed a chaincode-based NFT standard on \textsc{Fabric} according to the Ethereum \texttt{ERC-721} standard.

\section{Discussion on Mitigation Strategies}
        As a first step of the study, this work poses a quantitative analysis of the extent to which NFT is lost, inaccessible, and duplicated along the NFT-to-Asset connection. Technical enhancements and investors' awareness are needed to be made before the value of NFTs can be more permanently preserved and more widely recognized. To this end, we discuss mitigation strategies to resolve and alleviate the aforementioned issues. Potential solutions are suggested for NFT holders, developers and applications, respectively.
        
        \subsection{NFT (potential) holders.}
            For \textbf{NFT holders}, if the NFT is stored on centralized servers, we recommend they keep a copy of the data themselves and keep assets in separate locations to reduce the risk of data loss. Suppose the NFT is on a decentralized storage network. In that case, we recommend they recall the data regularly to keep NFT from being forgotten in the storage network and use a pinning service to enhance data durability. For \textbf{NFT potential holders}, they need to understand the potential risks of NFTs before investment and take a careful examination on: how the NFT is binding to the asset, on which platform it is hosting, whether there is hash value on-chain, whether the asset is unique, whether the NFT metadata in the smart contract can be modified at will, and whether the right to change is with the user or the application provider. 
        
        \subsection{NFT developers.}
            NFTs are in peril under the existing scheme, which poses new challenges to optimize the NFT mechanism to meet the security demands of current public NFTs and new emerging private NFTs \cite{scrt, arcana}. Notice that we will only discuss the solution for NFTs with off-chain storage, as on-chain storage is impractical for most of the scenarios due to the expensive on-chain storage cost. 
            In order to fully guarantee the equity of NFT holders, we first propose a definition of NFT safety, which is a \textbf{strong NFT-to-Asset association} mechanism that satisfies the following properties:
            
            \begin{itemize}
                \item \textbf{Tamper resistance.} \textit{The integrity and authenticity of NFT data can be publicly verified.} 
                \item \textbf{Sustainable storage.} \textit{Asset can be retrieved at most of the time.}
                \item \textbf{Traceable ownership.} \textit{The asset data can be traced back to the information on the blockchain for ownership verification}. 
            \end{itemize}
            
            Based on the above definition, we encourage \textbf{NFT developers} to mitigate the weaknesses of NFTs from the angles of:
            
            \ding{182} \textbf{Enhancement on the NFT-to-Asset connection.} 
            For public NFTs, the existing \textsc{IPFS} scheme meets the ownership traceability and tamper-resistance properties for public NFT data, as the data hash (that is the \texttt{CID} on \textsc{IPFS}) is recorded on-chain and forms an effective constraints to off-chain asset. Since the NFT is public, the asset can naturally be seen and owned by anyone. Data published on \textsc{IPFS} is publicly retrieval, unless it is encrypted in advance. Therefore, the \textsc{IPFS} solution towards traceable ownership is not applicable to private NFT data. 
            
            For private NFTs, it means that only the owners (past and current) of the NFT owns the digital asset. The record of asset hash enables an owner to self-prove the ownership of the off-chain data. In the current scheme, NFT sellers and buyers have exactly the same data before and after the transaction. As in the process, the blockchain will only change the NFT's ownership pointer from the seller to the buyer, and the asset will not take any change at all (so owners have a same asset data). When there occurs an illegal use of data and ownership determination is required, no available scheme is sufficient to trace back the breach as far as we know, as it is unable to determine which owner illegally leaks the data. We suggest to start from distinguishing data owned by different owners, and verify such different data by on-chain fuzzy hash. The different data has unique identification on the blockchain, and can thus be traced back to the ownership on the blockchain even when the muti-hop URL is broken.
            
            \ding{183} \textbf{Improvement on sustainability and availability of decentralized storage.}
            We use \textsc{IPFS} to represent decentralized storage based on its dominant position in decentralized storage. The \textsc{IPFS} data collection in this work shows great difference in accessibility and response duration when querying different gateways (nodes). Not only did our data access run into such problems, but also there are  complaints from the community, that files that have been uploaded and pinned to \textsc{IPFS} are unable to access. We notice that some works have explored the \textsc{IPFS} network, such as the characteristics of the underlying P2P network \cite{daniel2022passively, trautwein2022design}, the malicious resource occupancy attack \cite{patsakis2019hydras}, \textit{etc}. However, research has not been conducted on the causes of the aforementioned problem, such as the distribution of content replica (Replication Protocol), the efficiency of node discovery (\texttt{DHT}), the efficiency of content exchange (\textsc{Bitswap} Protocol), \textit{etc}., which leaves a barrier from better understanding the underlying reasons for the \textsc{IPFS}’s insufficient performance. We suggest the first step is to conduct a measurement of the effectiveness of the underlying protocol to identify the root cause. On this basis, the second step is to study and improve the bottleneck of each \textsc{IPFS} components to increase the performance. 
        
        \subsection{NFT applications.}
            Users(holders) are at significant risk because most of them have no actual control over their assets. We call on \textbf{NFT applications} to make a clear division of storage responsibility and the right of use with NFT holders. It prevents the problem of ambiguity of user and provider's responsibility when NFTs are lost in some unavoidable situations. We also recommend NFT applications to open-source their smart contracts and scrutinize the security via a trusted third party. Such security review will convince users of the authentication of the NFT structure, \textit{e.g.}, how the metadata is generated and fed on-chain, what features of the NFT are included in the metadata, and who can modify the metadata. At last, they can actively upgrade the technical framework and make efforts to integrate new technologies to improve the reliability of NFTs.

\section{Conclusion}
    As most of the corresponding assets of NFTs are stored off blockchain and NFTs bind on-chain identifiers with off-chain assets by multi-stage URLs, the weak binding easily leads to the disconnection between NFTs and their assets, resulting in the loss of the NFT value. 
    To this end, we have analyzed the phases along the path of NFT-to-Asset connection in detail, including the storage distribution, accessibility, and the degree of duplication. 
    Our results have shown that non-trivial portion of NFTs have been disconnected from their assets and have highly repetitive data within their collections, which significantly undermines the NFTs' value. We have further discovered that decentralized storage does not show advantages in accessibility, response time, and anti-loss for NFTs, which still needs to be strengthened.

\bibliographystyle{ACM-Reference-Format}
\bibliography{reference}

\end{document}